\documentclass[journal, twocolumn, 10pt, romanappendices,A4paper]{IEEEtran}
\usepackage{indentfirst}
\usepackage{setspace}
\usepackage[numbers,sort&compress]{natbib}
\usepackage{graphicx}
\usepackage{amstext}
\usepackage{algorithm}
\usepackage{algorithmic}
\usepackage{mathrsfs}
\usepackage{amssymb}
\usepackage{amsmath}
\usepackage{epstopdf}
\usepackage{CJK}
\usepackage{multicol}
\usepackage{stfloats}
\usepackage{url}
\usepackage{color}
\usepackage{multirow}
\usepackage{tabularx}
\usepackage{array}

\usepackage{graphicx,color,overpic,psfrag}
\usepackage{times}
\usepackage{latexsym}
\usepackage{bm}
\usepackage{amssymb}
\usepackage{cases}
\usepackage{array}
\usepackage{float}
\usepackage{fancyhdr}
\usepackage{booktabs}

\usepackage{psfrag}
\usepackage{colortbl}
\usepackage[table]{xcolor}
\usepackage{tabu}

\definecolor{mygray}{gray}{.9}
\definecolor{mypink}{rgb}{.99,.91,.95}
\definecolor{mycyan}{cmyk}{.3,0,0,0}

\newcommand{\PreserveBackslash}[1]{\let\temp=\\#1\let\\=\temp}
\newcolumntype{C}[1]{>{\PreserveBackslash\centering}p{#1}}
\newcolumntype{R}[1]{>{\PreserveBackslash\raggedleft}p{#1}}
\newcolumntype{L}[1]{>{\PreserveBackslash\raggedright}p{#1}}

\ifCLASSINFOpdf

\else

\fi

\begin{document}

\title{{On the Uplink Transmission of Extra-large Scale Massive MIMO Systems}}

\author{Xi Yang,~\IEEEmembership{Student Member,~IEEE}, Fan Cao,~\IEEEmembership{Student Member,~IEEE},\\ Michail Matthaiou,~\IEEEmembership{Senior Member,~IEEE}, and Shi Jin,~\IEEEmembership{Senior Member,~IEEE}

\thanks{Copyright (c) 2015 IEEE. Personal use of this material is permitted. However, permission to use this material for any other purposes must be obtained from the IEEE by sending a request to pubs-permissions@ieee.org.}
\thanks{Manuscript received February 23, 2020; revised June 16, 2020, August 8, 2020, and October 21, 2020; accepted November 4, 2020. This work was supported by the National Key Research and Development Program 2018YFA0701602, the National Science Foundation of China (NSFC) for Distinguished Young Scholars with Grant 61625106, the NSFC under Grant 61921004, and Shenzhen Basic Research JCYJ20170412104656685. The work of M. Matthaiou was supported by a research grant from the Department for the Economy Northern Ireland under the US-Ireland R\&D Partnership Programme and by the EPSRC, U.K., under Grant EP/P000673/1. The associate editor coordinating the review of this article and approving it for publication was Y. Zhou. (\emph{Corresponding author: Shi Jin.})}
\thanks{X. Yang, F. Cao, and S. Jin are with the National Mobile Communications Research Laboratory, Southeast University, Nanjing, 210096, P. R. China (e-mail: ouyangxi@seu.edu.cn; cao.fan@seu.edu.cn; jinshi@seu.edu.cn).}
\thanks{M. Matthaiou is with the Institute of Electronics, Communications and Information Technology (ECIT), Queen's University Belfast, Belfast, U.K. (e-mail: m.matthaiou@qub.ac.uk).}

}
\newcommand{\rl}[1]{\color{red}#1}

\maketitle

\begin{abstract}
With the inherent benefits, such as, better cell coverage and higher area throughput, extra-large scale massive multiple-input multiple-output (MIMO) has great potential to be one of the key technologies for the next generation wireless communication systems.
However, in practice, when the antenna dimensions grow large, spatial non-stationarities occur and users will only see a portion of the base station antenna array, which we call visibility region (VR). To assess the impact of spatial non-stationarities, in this paper, we investigate the uplink transmission of extra-large scale massive MIMO systems by considering VRs.
In particular, we first propose a subarray-based system architecture for extra-large scale massive MIMO systems. Then, tight closed-form uplink spectral efficiency (SE) approximations with linear receivers are derived.
With the objective of maximizing the achievable SE, we also propose schemes for the subarray phase coefficient design.
In addition, based on the obtained ergodic achievable SE approximations, two statistical channel state information (CSI)-based greedy user scheduling algorithms are developed.
{Our results indicate that the statistical CSI-based greedy joint user and subarray scheduling algorithm collaborating with the on-off switch-based subarray architecture is a promising practical solution for extra-large scale massive MIMO systems.}

\end{abstract}


%
\IEEEpeerreviewmaketitle
\newtheorem{Definition}{Definition}
\newtheorem{Lemma}{Lemma}
\newtheorem{Theorem}{Theorem}
\newtheorem{Corollary}{Corollary}
\newtheorem{Proposition}{Proposition}
\newtheorem{Remark}{Remark}

\section{Introduction}

%
%
%
%

Massive multiple-input multiple-output (MIMO), whose idea is to employ a large-scale antenna array at a base station (BS) to serve multiple users simultaneously, thus, achieving great spatial multiplexing gains and better spectral efficiency, has been identified as one of the key technologies in fifth-generation wireless communication systems \cite{Rusek13Scaling,Larsson14massive,Marzetta10noncooperative,Liu19Economically,Garcia14Coordinated,Liu18Load}.
When the antenna dimension continues to increase, the arrays turn out to be physically very large and the benefits such as, channel hardening, asymptotic inter-user channel orthogonality, cell coverage, area throughput, etc., promised by massive MIMO can be fully harnessed from a theoretical point of view. These extra-large scale antenna arrays could be developed and be integrated into large infrastructures, such as, the roof of airports, the walls of stadiums, or large shopping malls \cite{Martinez14towards}. With these enhanced benefits, extra-large scale massive MIMO is a very promising technology for the sixth-generation of wireless communications \cite{Emil19massive}.

However, the reality is not so idealistic: based on some recent measurement results in \cite{Gao15massive,Gao15Massiveallantenna}, when the antenna dimensions become large, spatial non-stationarities start to kick in. This arises from the fact that when the dimension of an antenna array is large, the far-field propagation assumption breaks down since the distances between the BS and scatterers or users are smaller than the Rayleigh distance.\footnote{The exterior fields of a radiating antenna (array) can be generally divided into two fields \cite{Yaghjian86overview}, i.e., the near field and the far field. The Rayleigh distance, which is equal to $(2{D}^2/\lambda+\lambda)$, is the dividing line between the near field and the far field, where $\lambda$ is the carrier wavelength and ${D}$ denotes the maximum electrical size of the antenna aperture. Note that the near-field region can be further divided into two subregions, i.e., the reactive near-field region and the radiating near-field region. The reactive near field is within a wavelength of the antenna (array), while the radiating near field is measured from the outer boundary of the reactive near field of the antenna (array) to the Rayleigh distance. In this paper, we mainly focus on the communication in the radiating near field.}
{Thus, in contrast to traditional massive MIMO systems equipped with compact uniform planar arrays, extra-large scale massive MIMO systems with extra-large antenna array size (for example, massive antennas coated on the ceiling of a vast football stadium or the walls of a hundred-meter tall building \cite{Ericsson19}) generally operate in near-field propagation conditions.
In these extra-large scale massive MIMO systems, users can only see a portion of the BS antenna array due to the rapid power attenuation of the transmitted signals and the extra-large array size.} The portion of the antenna array at the BS seen by users is called visibility region (VR) \cite{Carvalho19nonstationary}. Each user has its specific VR and the locations of VRs for different users can be separate, partially overlapped, or completely overlapped, depending on the surrounding environment and the users' relative positions along the antenna array.
{Note that the difference between the traditional stationary massive MIMO and the extra-large massive MIMO is determined by the relationship between users' location and Rayleigh distance with respect to the maximum antenna array size, instead of merely the number of antennas.}

Due to the existence of VRs, the performance characterization of extra-large scale massive MIMO systems is different from that of stationary massive MIMO systems by simply letting their number of BS antennas go to infinity.
There are a few works exploiting the performance of extra-large scale massive MIMO systems. In \cite{Amiri18extremely}, with the objective of improving the computational efficiency, a disjoint subarray-based receiver architecture and distributed linear data fusion receiver with bipartite graph-based user selection method were proposed. The work of \cite{Li15capacity} presented an ergodic capacity analysis of extra-large scale massive MIMO by introducing a tractable non-stationary channel model which divides the scattering clusters into two categories, i.e., wholly visible clusters and partially visible clusters, and regards these clusters as an array with virtual antennas.
A simple non-stationary channel model, which has connections with the stationary massive MIMO channel model, was proposed in \cite{Ali19linear}. {The downlink analysis of extra-large scale massive MIMO systems with linear precoders was also provided by deriving an approximate deterministic equivalent of the instantaneous signal-to-interference-plus-noise ratio (SINR).} The numerical results in \cite{Ali19linear} indicated that the VR significantly impacts the performance of linear precoders.
Despite that, there is less of work that investigates the uplink performance of extra-large scale massive MIMO systems when taking the spatial non-stationarities into account.

Motivated by these existing works, we mainly focus on the uplink transmission of  extra-large scale massive MIMO systems by considering VRs. In particular, we firstly propose a practical system architecture suitable for extra-large scale massive MIMO, where a subarray-based hybrid architecture is adopted to alleviate the overall hardware cost and complexity. Then, the uplink achievable spectral efficiencies (SEs) of the extra-large scale massive MIMO system with linear receivers are examined. Afterwards, we investigate the design of the subarray in order to maximize the achievable SE. Two statistical channel state information (CSI)-based greedy user scheduling algorithms are also proposed and numerical simulations are performed to validate their performance.
The main contributions of this paper can now be summarized as follows:
\begin{itemize}
\item We derive tight closed-form ergodic uplink achievable SE approximations for the  extra-large scale massive MIMO system with linear receivers, i.e., maximum ratio combining (MRC) receiver and linear minimum mean squared error (LMMSE) receiver. The ergodic achievable SE approximation for the MRC receiver shows that, in order to maximize the system sum achievable SE, users with their VRs covering different subarrays or VRs with less overlap should be simultaneously scheduled. On the other hand, the ergodic achievable SE approximation for the LMMSE receiver indicates that we should simultaneously schedule as many users as possible.
\item By considering two subarray architectures, i.e., the subarray with phase shifters and the on-off switch-based subarray, we investigate the design of the subarray for the extra-large scale massive MIMO system. {For the subarray with phase shifters, to boost the sum achievable SE, the optimal phase coefficients are the phases of the eigenvectors corresponding to the maximum eigenvalues of the main block matrices of the channel correlation matrix. For the on-off switch-based subarray, the user who has the larger sum energy radiated to the subarrays should be selected for communication.}
\item Based on the obtained ergodic achievable SE approximations, we propose two statistical CSI-based greedy scheduling algorithms, i.e., the statistical CSI-based greedy user scheduling algorithm and the statistical CSI-based greedy joint user and subarray scheduling algorithm, for the purpose of maximizing the system achievable SE. {Numerical results manifest that in the extra-large scale massive MIMO regime, it is not necessary to simultaneously turn on all subarrays and radio frequency (RF) chains to serve the users at uplink. The introduction of dynamic subarray scheduling is beneficial to achieve better system performance with lower energy consumption, and the on-off switch-based subarray architecture with LMMSE receiver is a promising low cost solution, especially in terms of system energy efficiency.}
\end{itemize}

The rest of this paper is organized as follows: In Section II, we present the system architecture and the signal model for the extra-large scale massive MIMO system. Section III investigates the uplink ergodic achievable SEs under the linear receivers and Section IV provides the phase coefficient design of subarrays. The proposed statistical CSI-based user scheduling algorithms are provided in Section V. Section VI presents the numerical results and we conclude the paper in Section VII.

Throughout the paper, we use bold lowercase $\mathbf{a}$ and bold uppercase $\mathbf{A}$ to denote vectors and matrices, respectively. The superscripts $(\cdot)^{*}$, $(\cdot)^{T}$, and $(\cdot)^{H}$ represent the conjugate, transpose, and conjugate-transpose operations of matrix, respectively; $\mathbf{I}_N$ is an identity matrix with dimension $N\times N$; $\odot$ and $\otimes$ denote the element-wise product and the Kronecker product, respectively. Also, $\mathbb{E}\{\cdot\}$ is the expectation operation, $\left\| \mathbf{a} \right\|$ stands for the norm of the vector $\mathbf{a}$; ${\rm tr}(\mathbf{A})$, $\det(\mathbf{A})$, and ${\mathbf{A}^{-1}}$ stand for the trace, determinant, and inverse of the matrix $\mathbf{A}$, respectively. $[\mathbf{A}]_{ij}$ denotes the element at the $i$th row and the $j$th column of $\mathbf{A}$. ${\rm diag}(x_1,x_2,...,x_N)$ represents a diagonal matrix with diagonal elements $x_i,\;i=1,\ldots,N$, while ${\rm{blkdiag}}({{\bf{X}}_1},{{\bf{X}}_2}, \ldots ,{{\bf{X}}_N})$ represents a diagonal matrix with block diagonal matrices ${{\bf{X}}_i},\;i=1,\ldots,N$.

\section{System Model}
In this section, we firstly describe the system architecture of the extra-large scale massive MIMO system, and then the signal model is provided.
\subsection{System Architecture}
\begin{figure*}[!t]
	\centering
	\includegraphics[scale= 0.5]{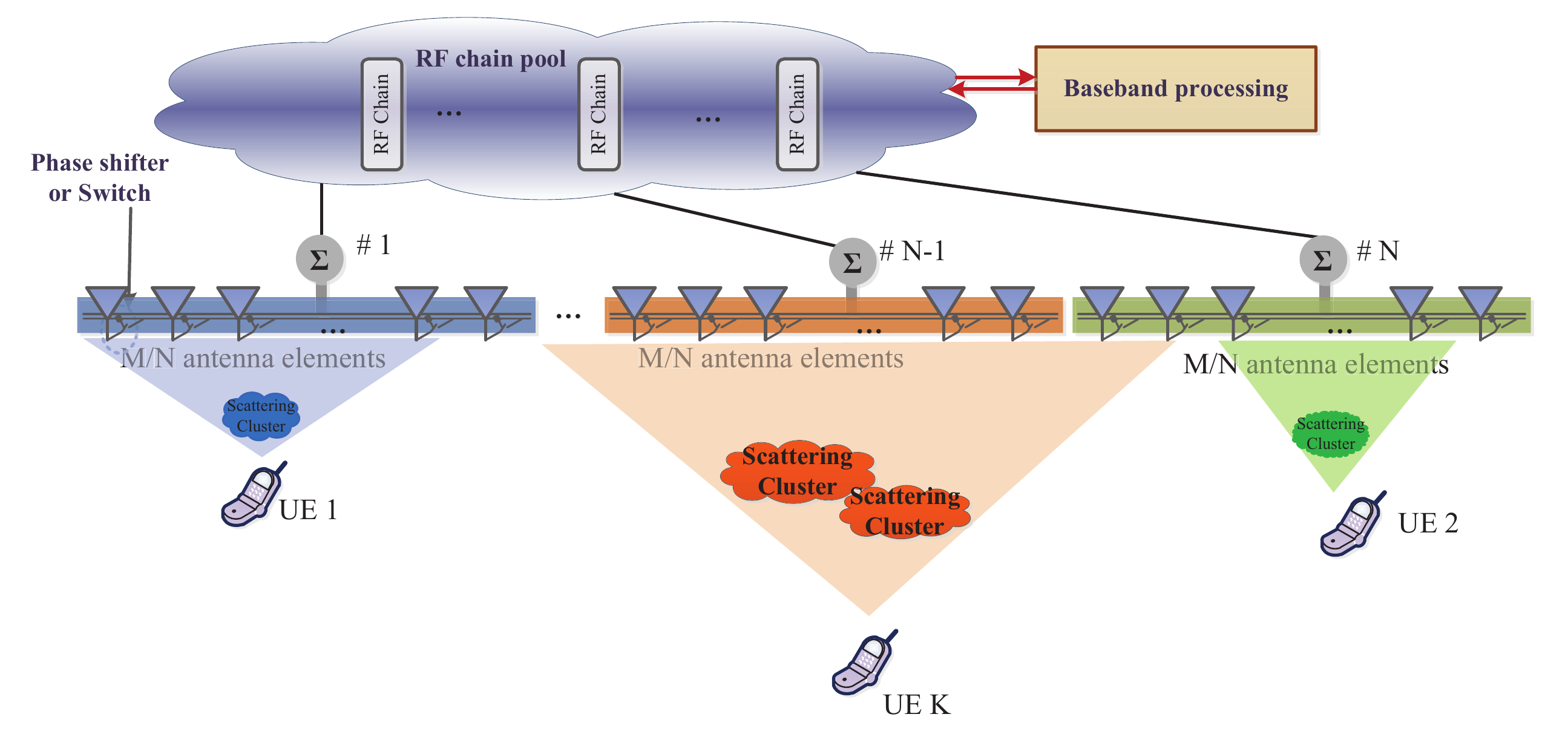}
	\caption{The system architecture of an extra-large scale massive MIMO system, in which a BS equipped with an $M$-element large uniform linear array (ULA) serves $K$ single-antenna user simultaneously. A subarray-based hybrid architecture is adopted at BS. Each subarray includes $M/N$ antenna elements and there are $N$ subarrays in total.}\label{Fig:system_architecture}
\end{figure*}

Consider an extra-large scale massive MIMO system illustrated in Fig.\,\ref{Fig:system_architecture}, where a BS equipped with an $M$-element large uniform linear array (ULA)\footnote{In this paper, we take ULA as a simple example, yet, the proposed system architecture can also be applied to other antenna array topologies, such as, uniform planar antenna array (UPA), and so on. } serves $K$ single-antenna users simultaneously.

Since the massive number of antennas at the BS render an independent RF chain per antenna element impractical in terms of hardware cost and system complexity, we propose a subarray-based hybrid architecture\footnote{Hybrid analog and digital beamforming architectures, including the fully-connected architecture (i.e., each phase shifter is connected to all the BS antennas) and the subarray-based architecture, have been widely applied in traditional massive MIMO systems \cite{Ayach14spatially,Bogale16On,Alkhateeb16Massive}. We mainly focus on the subarray-based hybrid beamforming architecture in this paper due to the extremely large size antenna array in extra-large scale massive MIMO systems.} as presented in Fig.\,\ref{Fig:system_architecture}. The BS consists of subarrays, an RF chain pool, and a baseband processing unit. Each subarray includes $M/N$ antenna elements and, thus, $N$ subarrays are configured. The RF chain pool contains RF chains and each RF chain can be statically or dynamically assigned to a dedicated subarray. Digital processing, such as, channel estimation, data detection, or user scheduling is performed in the baseband processing unit.
Note that each subarray is connected with an RF chain and therefore can support one data stream. Due to existence of VR in extra-large scale massive MIMO systems, multiple consecutive subarrays could be covered by one user. Moreover, when overlapped VRs occur, the multiple consecutive subarrays may also support other users simultaneously, which inevitably creates inter-user interference.

To harvest the array gain provided by the large number of antennas, two different subarray architectures i.e., the subarray with phase shifters and the on-off switch-based subarray, are considered. {In the former architecture, each antenna element in the subarray is connected with a phase shifter, while in the latter type, the phase shifter is replaced by a switch.}
Hence, the signals acquired by antennas in a on-off switch-based subarray are directly combined without adjustable phase shifts before conveyed to the RF chain in the uplink. It is worth noting that compared with the subarray with phase shifters and despite the anticipated performance loss, the on-off switch-based subarray is much cheaper and more hardware-implementation friendly, especially in extra-large scale massive MIMO regime. In addition, as will be presented in the numerical results, the on-off switch-based subarray also yields great performance when combined with the LMMSE receiver.

\subsection{Signal Model}

We focus on the uplink transmission of the  extra-large scale massive MIMO system in this paper. Taking the spatial non-stationarity of the extra-large scale MIMO system channels into consideration, we model the channel ${{\bf{h}}_k} \in {\mathbb{C}^{M \times 1}}$ between the $k$-th user and the BS as\footnote{Although the channel model in (\ref{1}) is usually employed for sub-6G wireless channel modeling, the subsequent analysis in Section III is actually irrelevant to the specific structure of ${{\bf{\Theta }}_k}$. On the other hand, the finite-dimensional channel model \cite{BradyBeamspace2013,Sayeed2002}, which is typically used for millimeter wave band channel modeling, can also be casted into a similar expression, i.e., ${{\bf{h}}_k} = {\bf{\tilde A}}_k{{\bf{g}}_k}$, where ${\bf{\tilde A}}_k={\bf{D}}_k{\bf{A}}_k$, ${\bf{A}}_k$ contains the steering vectors, and ${{\bf{g}}_k} \sim {\cal C}{\cal N}\left( {{\bf{0}},{{\bf{I}}_P}} \right)$ contains the complex path gains. Hence, the analysis presented in Section III can also be applied to millimeter wave systems.}
\begin{equation}\label{1}
{{\bf{h}}_k} = {\bf{\Theta }}_k^{1/2}{{\bf{g}}_k},
\end{equation}
where ${{\bf{g}}_k} \sim {\cal C}{\cal N}\left( {{\bf{0}},{{\bf{I}}_M}} \right)$ and ${{\bf{\Theta }}_k}\in {\mathbb{C}^{M \times M}}$ represents the correlation matrix at the BS for the $k$-th user, given as \cite{Ali19linear}
\begin{equation}
{{\bf{\Theta }}_k} = {\bf{D}}_k^{1/2}{{\bf{R}}_k}{\bf{D}}_k^{1/2}.
\end{equation}
Note that ${{\bf{R}}_k}\in {\mathbb{C}^{M \times M}}$ denotes the spatial correlation matrix of user $k$ corresponding to the case of a stationary massive MIMO channel, whose structure relies on the antenna array topology at the BS and the angular spread. The spatial non-stationarity, i.e, the user $k$'s VR, in the extra-large scale massive MIMO channel is characterized by the real diagonal matrix
\begin{equation}
{{\bf{D}}_k} = {\rm{diag}}(d_1^{(k)},d_2^{(k)}, \ldots ,d_M^{(k)}),
\end{equation}
where $d_m^{(k)}, m=1,\ldots,M,$ denotes the spatial non-stationarity for user $k$ at the antenna element $m$. {Note that the large scale fading, e.g., the path loss, of user $k$ can be still incorporated in $d_m^{(k)}$.} We point out only a few diagonal elements of ${{\bf{D}}_k}$ are non-zero \cite{Ali19linear},\footnote{Although the subsequent analysis is applicable for general cases (e.g., the amplitude distribution of the non-zero diagonal elements forms a Hamming window or complies with the spherical wave propagation), we set the non-zero diagonal elements of ${{\bf{D}}_k}$ to be $1$ in the simulations section for simplicity. } and the number of non-zero diagonal elements directly reflects the size of the user's VR. Different users can have different VR sizes, owing to their distances to the BS or their different uplink transmit power. When a ULA is employed at BS, the ${{\bf{R}}_k}$ in ${\bf{\Theta }}_k$ can be expressed as \cite{McKay06Random}
\begin{equation}\label{2}
{{\bf{R}}_k} = \left[ {{\bf{a}}({\theta _k}){{\bf{a}}^H}({\theta _k})} \right] \odot {\bf{P}}({\theta _k},{\sigma _k}),
\end{equation}
where ${\bf{a}}({\theta _k})\in {\mathbb{C}^{M \times 1}}$ is the steering vector of the ULA, defined as
\begin{equation}\label{3}
{\bf{a}}({\theta _k}) = {[1,{e^{j2\pi d\sin {\theta _k}}}, \ldots ,{e^{j2\pi (M - 1)d\sin {\theta _k}}}]^T},
\end{equation}
where ${\theta _k}$ represents the mean angle of arrival (AoA) of the $k$-th user and $d$ denotes the antenna element spacing normalized by the carrier wavelength. In our simulations, we set $d=1/2$.
Most importantly, ${\bf{P}}({\theta _k},{\sigma _k})\in {\mathbb{C}^{M \times M}}$ captures the angular spectrum of AoA and its entries come from a Gaussian angular spread distribution with variance $\sigma _k^2$ \cite{McKay06Random}. The ${\left\{ {m,n} \right\}^{th}}$ entry of ${\bf{P}}({\theta _k},{\sigma _k})$ can be given by
\begin{equation}\label{4}
{\{ {\bf{P}}({\theta _k},{\sigma _k})\} _{m,n}} = {e^{ - 2[\pi d{{(m - n)}^2}]\sigma _k^2{{\cos }^2}{\theta _k}}},\quad m,n = 1, \ldots ,M.
\end{equation}

Therefore, in the uplink transmission of the  extra-large scale MIMO system, the received signal at the BS can be written as
\begin{equation}\label{5}
{\bf{y}} = \sqrt {{p_u}} {\bf{Hx}} + {\bf{n}},
\end{equation}
where ${{p_u}}$ is the transmit power of each user, ${\bf{H}} = [{{\bf{h}}_1},{{\bf{h}}_2}, \ldots ,{{\bf{h}}_K}]$ represents the multi-user uplink channel matrix, and ${\bf{x}} = {[{x_1},{x_2}, \ldots ,{x_K}]^T}$ is the transmitted signal from $K$ users with ${x_k} \sim {\cal C}{\cal N}(0,1)$ for $k=1,2,...,K$; ${\bf{n}}$ is the complex Gaussian noise satisfying ${\bf{n}} \sim {\cal C}{\cal N}\left( {{\bf{0}},{\sigma ^2}{\bf{I}}_M} \right)$. Without loss of generality, we set ${\sigma ^2} = 1$. Additionally, for ease of exposition, we assume that $\mathbb{E}\{ {{\bf{g}}_i}{\bf{g}}_k^H\}  = {\bf{0}},\forall i \ne k$, that is to say, there is no correlation between any pair of channels across different users.

Since we employ a subarray-based hybrid architecture in the  extra-large scale MIMO system as presented in Fig.\,\ref{Fig:system_architecture}, the received signals at BS will be firstly combined in the analog domain and then be linearly demodulated in the digital domain. Thus, the processing procedure can be formulated as
\begin{equation}\label{6}
{\bf{r}} = \sqrt {{p_u}} {{\bf{A}}^H}{{\bf{W}}^H}{\bf{Hx}} + {{\bf{A}}^H}{{\bf{W}}^H}{\bf{n}},
\end{equation}
where ${\bf{r}}\in {\mathbb{C}^{K \times 1}}$ is the recovered signal vector; ${\bf{W}} = {\rm{blkdiag}}({{\bf{w}}_1},{{\bf{w}}_2}, \ldots ,{{\bf{w}}_N}) \in\mathbb{C} {^{M \times N}}$ represents the combining matrix in the analog domain and
${{\bf{w}}_i} \in \mathbb{C}  {^{(M/N) \times 1}}$ is a constant modulus vector, that is to say, all the elements of ${{\bf{w}}_i}$ have constant amplitude of $\sqrt {N/M}$ and ${\bf{w}}_i^H{{\bf{w}}_i} = 1$ for $i=1,2,...,N$.\footnote{When the on-off switch-based subarrays are deployed, we assume that the number of the specified switches at ``ON'' status in each subarray is equal and the value of ${\bf{w}}_i^H{{\bf{w}}_i}$ should be corrected with a scalar according to the number of the switches at ``ON'' status.} Also, ${\bf{A}} \in \mathbb{C} {^{N \times K}}$ is the linear detection matrix in the digital domain, which has different expressions for different linear receivers such as, MRC receiver and LMMSE receiver. Here, we define ${\bf{F}} = {{\bf{W}}^H}{\bf{H}}$ and assume that perfect CSI is available at the BS\footnote{Channel state information can be obtained by various methods, such as, by sending orthogonal pilots from users. {Similar to \cite{Li15capacity} and \cite{Ali19linear}, we assume perfect CSI at the receiver in order to assess in detail the impact of non-stationarities without complicating the notation.}
}, then ${\bf{A}}$ can be given by
\begin{equation}\label{8}
\begin{aligned}
{\bf{A}} = \left\{ {\begin{array}{*{20}{c}}
	{{\bf{F}},                                                                     \quad   \quad \quad   \quad   \quad   \quad \quad  \quad {{\rm{for\quad MRC}}}},\\
	{{\bf{F}}{{\left( {{{\bf{F}}^H}{\bf{F}} + \frac{1}{{{p_u}}}{{\bf{I}}_K}} \right)}^{ - 1}},     \quad{{\rm{for\quad  LMMSE}}}}.
	\end{array}} \right.
\end{aligned}
\end{equation}
Hence, the signal of the $k$-th user at the BS can be expressed as
\begin{equation}\label{9}
{r_k} = \sqrt {{p_u}} {\bf{a}}_k^H{{\bf{W}}^H}{{\bf{h}}_k}{x_k} + \sqrt {{p_u}} \sum\limits_{i = 1,i \ne k}^K {{\bf{a}}_k^H{{\bf{W}}^H}{{\bf{h}}_i}{x_i}}  + {\bf{a}}_k^H{{\bf{W}}^H}{\bf{n}},
\end{equation}
where ${{\bf{a}}_k} = {\bf{A}}(:,k)$ is the $k$-th column of the matrix {\bf A}.

By modeling the noise-plus-interference term as additive Gaussian noise independent of ${x_k}$ with zero mean and variance of ${{p_u}\sum\limits_{i = 1,i \ne k}^K {|{\bf{a}}_k^H{{\bf{W}}^H}{{\bf{h}}_i}{|^2}}  + {{\left\| {{\bf{a}}_k^H{{\bf{W}}^H}} \right\|}^2}}$ \cite{Ngo13energy}, we obtain the ergodic achievable SE of the $k$-th user in the uplink data transmission as
\begin{align}\label{Rk_general}
&{R_k}= \nonumber\\& \mathbb{E}{_{\bf{h}}}\left\{ {{{\log }_2}\left( {1 + \frac{{{p_u}|{\bf{a}}_k^H{{\bf{W}}^H}{{\bf{h}}_k}{|^2}}}{{{p_u}\sum\limits_{i = 1,i \ne k}^K {|{\bf{a}}_k^H{{\bf{W}}^H}{{\bf{h}}_i}{|^2}}  + {{\left\| {{\bf{a}}_k^H{{\bf{W}}^H}} \right\|}^2}}}} \right)} \right\},
\end{align}
and the sum uplink achievable SE of the  extra-large scale massive MIMO system becomes
\begin{equation}\label{Rsum_general}
R = \sum\limits_{i = 1}^K {R{}_i}.
\end{equation}

In the next section, we aim to examine the ergodic achievable SEs in (\ref{Rk_general}) and (\ref{Rsum_general}) under two different types of receivers, i.e., MRC receiver and LMMSE receiver, and then identify the influence of specific VR distributions on the system ergodic achievable SE.

\vspace{-0.2cm}
\section{Uplink Achievable Spectral Efficiency Analysis}
In this section, we investigate the ergodic uplink achievable SEs of linear receivers, i.e., MRC receiver and LMMSE receiver, for extra-large scale massive MIMO systems.

\subsection{MRC Receiver}
When the MRC receiver is employed at the BS, the approximation of the ergodic achievable SE is provided in the following theorem.
\begin{Theorem}\label{Th1}
When the MRC receiver is adopted in the uplink of the  extra-large scale massive MIMO system, the ergodic achievable SE of the $k$th user can be approximated by
\begin{align}\label{Rk_MRC_app_k}
&R_k^{\text{MRC,app}} = \nonumber\\& {\log _2}\left( {1 + \frac{{{\rm{tr}}({\mathbf{B}}{{\mathbf{\Theta }}_k}{\mathbf{B}}{{\mathbf{\Theta }}_k}) + {\rm{t}}{{\rm{r}}^2}({\mathbf{B}}{{\mathbf{\Theta }}_k})}}{{\sum\limits_{i = 1,i \ne k}^K {{\rm{tr}}({\mathbf{B}}{{\mathbf{\Theta }}_i}{\mathbf{B}}{{\mathbf{\Theta }}_k})}  + \frac{1}{{{p_u}}}{\rm{tr}}({\mathbf{B}}{{\mathbf{B}}^H}{{\mathbf{\Theta }}_k})}}} \right),
\end{align}
where ${\mathbf{B}} \triangleq {\mathbf{W}}{{\mathbf{W}}^H}$.
\end{Theorem}
\begin{IEEEproof}
When the MRC receiver is employed at the BS, we have the linear detection matrix ${\bf{A}} = {\bf{F}}$, and
\begin{equation}\label{ak_MRC}
{{\bf{a}}_k} = {{\bf{W}}^H}{{\bf{h}}_k}.
\end{equation}
Substituting (\ref{ak_MRC}) into (\ref{Rk_general}), the ergodic uplink achievable SE for the $k$th user can be written as
\begin{align}\label{Rk_MRC}
  &R_k^{\text{MRC}}\nonumber\\
       &\mathop  \approx \limits^{(a)} {\log _2}\left( {1 + \frac{{{p_u}{\mathbb{E}_{\mathbf{h}}}\{ |{\mathbf{h}}_k^H{\mathbf{B}}{{\mathbf{h}}_k}{|^2}\} }}{{{p_u}\sum\limits_{i = 1,i \ne k}^K {{\mathbb{E}_{\mathbf{h}}}\{ |{\mathbf{h}}_k^H{\mathbf{B}}{{\mathbf{h}}_i}{|^2}\} }  + {\mathbb{E}_{\mathbf{h}}}\left\{ {{{\left\| {{\mathbf{h}}_k^H{\mathbf{B}}} \right\|}^2}} \right\}}}} \right),
\end{align}
where (a) applies the approximation $\mathbb{E}\{ {\log _2}(1 + X/Y)\}\approx {\log _2}(1 + \mathbb{E}\{ X\} /\mathbb{E}\{ Y\} )$ from \cite{Zhang14power}. Note that ${\mathbf{W}}$ is a block diagonal matrix owing to the subarray-based hardware architecture, thus ${\mathbf{B}}= {\mathbf{W}}{{\mathbf{W}}^H}$ is also a block diagonal matrix.
Therefore, the results in (\ref{Rk_MRC_app_k}) can be derived directly by calculating the terms in the numerator and the denominator of (\ref{Rk_MRC}) as following:
\begin{align}\label{T1.1}
{\mathbb{E}_{\mathbf{h}}}\{ {\mathbf{h}}_k^H{\mathbf{B}}{{\mathbf{h}}_k}\}
&= {\rm{tr}}({\mathbf{B}}{{\mathbf{\Theta }}_k}),
\end{align}
\begin{equation}\label{T1.2}
{\mathbb{E}_{\mathbf{h}}}\{ |{\mathbf{h}}_k^H{\mathbf{B}}{{\mathbf{h}}_k}{|^2}\}  = {\rm{tr}}({\mathbf{B}}{{\mathbf{\Theta }}_k}{\mathbf{B}}{{\mathbf{\Theta }}_k}) + {\rm{t}}{{\rm{r}}^2}({\mathbf{B}}{{\mathbf{\Theta }}_k}),
\end{equation}
\begin{align}\label{T1.3}
{\mathbb{E}_{\mathbf{h}}}\{ |{\mathbf{h}}_k^H{\mathbf{B}}{{\mathbf{h}}_i}{|^2}\}
                           &= {\rm{tr}}({\mathbf{B}}{{\mathbf{\Theta }}_i}{\mathbf{B}}{{\mathbf{\Theta }}_k}),
\end{align}
\begin{equation}\label{T1.4}
{\mathbb{E}_{\mathbf{h}}}\left\{ {{{\left\| {{\mathbf{h}}_k^H{\mathbf{B}}} \right\|}^2}} \right\} = {\rm{tr}}({\mathbf{B}}{{\mathbf{B}}^H}{{\mathbf{\Theta }}_k}).
\end{equation}
Substituting (\ref{T1.1})-(\ref{T1.4}) into (\ref{Rk_MRC}), we obtain (\ref{Rk_MRC_app_k}).
\end{IEEEproof}

It is worth to mention that the derivations in Theorem\,\ref{Th1} do not rely on the specific structure of ${\mathbf{\Theta }_k}$ in terms of the antenna array topology, thus, the results in Theorem\,\ref{Th1} can also be applied to other antenna array topologies e.g., UPA.
Observe from Theorem\,\ref{Th1} that, the ergodic achievable SE is dominated by the matrix product ${\mathbf{B}}{\mathbf{\Theta }}$, which therefore necessitates the design of subarray corresponding to the users' correlation matrices. In addition, the impact of VR characterized by ${\bf{D}}$ is contained in ${\mathbf{\Theta }}$, which finally constrains the increase of the ergodic achievable SE with the number of antennas at BS, even if there is no multi-user interference. For example, for user $k$ we assume
\begin{equation}
{{\mathbf{D}}_k} = {\text{diag}}( 0,\ldots ,0,\underbrace {d_{i + 1}^{(k)},d_{i + 2}^{(k)}, \ldots ,d_{i + E}^{(k)}}_{{\text{VR of user }}k},0, \ldots,0 ),\nonumber
\end{equation}
where $E$ denotes the number of antennas covered by the VR of user $k$. Then, only these $E$ antennas at the BS actually collect the signals radiated by the user $k$. In other words, no matter how many antennas the BS is equipped with, the number of effective receive antennas is mainly limited by $E$. Thereby, the increase of user $k$'s uplink ergodic achievable SE is severely compromised.

What is more, to maximize the ergodic achievable SE per user and consequently maximize the system sum achievable SE, the multi-user interference $\sum\limits_{i = 1,i \ne k}^K {{\rm{tr}}({\mathbf{B}}{{\mathbf{\Theta }}_i}{\mathbf{B}}{{\mathbf{\Theta }}_k})}$ term should be minimized in (\ref{Rk_MRC_app_k}).
Assume now that
\begin{equation}
{\mathbf{B}} = \rm{blkdiag}({{\mathbf{\bar B}}_1},{{\mathbf{\bar B}}_2}, \ldots ,{{\mathbf{\bar B}}_N}),
\end{equation}
then
\begin{equation}
{{\mathbf{\bar B}}_i} = {{\mathbf{w}}_i}{\mathbf{w}}_i^H,i = 1, \ldots ,N.
\end{equation}
Since ${\rm{tr}}({\mathbf{B}}{{\mathbf{\Theta }}_i}{\mathbf{B}}{{\mathbf{\Theta }}_k}) \geqslant 0$, ${{\mathbf{\Theta }}_i} = {\mathbf{D}}_i^{1/2}{{\mathbf{R}}_i}{\mathbf{D}}_i^{1/2}$ and ${{\mathbf{\Theta }}_k} = {\mathbf{D}}_k^{1/2}{{\mathbf{R}}_k}{\mathbf{D}}_k^{1/2}$ are block diagonal matrices, and ${\mathbf{B}}$ is a block diagonal matrix corresponding to the subarray architecture as well, we can obtain ${\rm{tr}}({\mathbf{B}}{{\mathbf{\Theta }}_i}{\mathbf{B}}{{\mathbf{\Theta }}_k}) = 0$ when ${\mathbf{1}}({{\mathbf{D}}_i}) \odot {\mathbf{1}}({{\mathbf{D}}_k}) = {\mathbf{0}}$, in which ${\mathbf{1}}({{\mathbf{D}}_i})$ denotes the $N$-dimension indicator function with its $n$th element calculated by
\begin{equation}\label{T1.5}
{[{\mathbf{1}}({{\mathbf{D}}_i})]_n} = \left\{ {\begin{array}{*{20}{c}}
  {1,\quad {{\mathbf{D}}_i} \odot {\rm{blkdiag}}({{\mathbf{0}}},\ldots ,{\bar{\mathbf{B}}_n}, \ldots ,{{\mathbf{0}}}) \ne {\mathbf{0}}}, \\
  {0,\quad {{\mathbf{D}}_i} \odot {\rm{blkdiag}}({{\mathbf{0}}},\ldots ,{\bar{\mathbf{B}}_n}, \ldots ,{{\mathbf{0}}}) = {\mathbf{0}}}.
\end{array}} \right.
\end{equation}
As a result, we obtain the following remark.
\begin{Remark}\label{Remark1}
For the MRC receiver, in order to maximize the system sum achievable SE, users with their VRs covering different subarrays or VRs with less overlap should be scheduled simultaneously.
\end{Remark}

Furthermore, by recalling that the MRC receiver has no capability of cancelling inter-user interference, which becomes more problematic in the high signal-to-noise ratio (SNR) regime, we exploit the LMMSE receiver in the next subsection.

\subsection{LMMSE Receiver}
When the LMMSE receiver is employed at the BS, we have the linear detection matrix ${\mathbf{A}} = {\mathbf{F}}{\left( {{{\mathbf{F}}^H}{\mathbf{F}} + \frac{1}{{{p_u}}}{{\mathbf{I}}_K}} \right)^{ - 1}}$. Substituting ${\mathbf{A}}$ into (\ref{Rk_general}), the ergodic achievable SE of the $k$th user under LMMSE receiver can be expressed as
\begin{align}\label{Rk_LMMSE1}
R_k^{\text{LMMSE}} &= {\mathbb{E}_{\mathbf{h}}}\left\{ {{{\log }_2}\left( {\frac{1}{{{{\left[ {{{\left( {{{\mathbf{I}}_K} + {p_u}{{\mathbf{F}}^H}{\mathbf{F}}} \right)}^{ - 1}}} \right]}_{kk}}}}} \right)} \right\}.
\end{align}
Since ${[{{\mathbf{M}}^{ - 1}}]_{kk}} = {{\det ({\mathbf{M}^{kk}}}) \mathord{\left/ {\vphantom {{\det ({\mathbf{M}})} {\det ({{\mathbf{M}}^{kk}})}}} \right. \kern-\nulldelimiterspace} {\det ({{\mathbf{M}}})}}$ where ${{{\mathbf{M}}^{kk}}}$ is the $(k,k)$th minor of the matrix ${\mathbf{M}}$ \cite{Horn90matrix}, combining ${({{\mathbf{F}}^H}{\mathbf{F}})^{kk}} = {\mathbf{F}}_{(k)}^H{{\mathbf{F}}_{(k)}}$ where ${\mathbf{F}}_{(k)}$ represents ${\mathbf{F}}$ with the $k$th column removed \cite{Matthew10achievable},
we can rewrite (\ref{Rk_LMMSE1}) as
\begin{align}\label{Rk_LMMSE}
R_k^{\text{LMMSE}}=& {\mathbb{E}_{\mathbf{h}}}\left\{ {{{\log }_2}\det \left( {{{\mathbf{I}}_K} + {p_u}{{\mathbf{F}}^H}{\mathbf{F}}} \right)} \right\} \nonumber \\&- {\mathbb{E}_{\mathbf{h}}}\left\{ {{{\log }_2}\det \left( {{{\mathbf{I}}_{K - 1}} + {p_u}{\mathbf{F}}_{(k)}^H{{\mathbf{F}}_{(k)}}} \right)} \right\}.
\end{align}
Note that it is greatly challenging, if not impossible, to directly evaluate (\ref{Rk_LMMSE}) under general cases, hence, in what follows we analyze the ergodic achievable SE by giving a separate treatment for two cases, i.e., (i) completely overlapped VR case, (ii) partially overlapped VR case. {The following analysis for these two cases mainly relies on their different characteristics originating from their different VR overlapping.
It is also worth mentioning that the scenario, where the VRs of different users do not overlap, is regarded as a special case of the partially overlapped one. }

\emph{\textbf{Completely Overlapped VR Case}}: In this case, users are closely distributed in a relatively small region in front of the extra-large scale massive MIMO system, therefore, the VRs of different users completely overlap. {Since the VRs completely overlap, the non-zero portions of ${\mathbf{\Theta }}$ for different users have the same location, and we assume that the size of VR is $\tilde M$.} To further simplify the problem, we also assume that ${{\mathbf{\Theta }}_1} =  \cdots  = {{\mathbf{\Theta }}_K} = {\mathbf{\Theta }}$, then
\begin{align}
{\mathbf{H}}
       &= {{\mathbf{\Theta }}^{1/2}}{\mathbf{G}},
\end{align}
where ${\mathbf{G}} \triangleq [{{\mathbf{g}}_1},{{\mathbf{g}}_2}, \ldots ,{{\mathbf{g}}_K}]$ and ${\mathbf{G}} \sim \mathcal{C}\mathcal{N}\left( {{\mathbf{0}},{{\mathbf{I}}_M} \otimes {{\mathbf{I}}_K}} \right)$. Therefore,
\begin{align}\label{FHF_complete_overlap}
{{\mathbf{F}}^H}{\mathbf{F}}
    &= {{\mathbf{G}}^H}{\mathbf{\tilde \Theta G}},
\end{align}
where we define ${\mathbf{\tilde \Theta }} \triangleq {{\mathbf{\Theta }}^{1/2}}{\mathbf{W}}{{\mathbf{W}}^H}{{\mathbf{\Theta }}^{1/2}}$.
Substituting (\ref{FHF_complete_overlap}) into (\ref{Rk_LMMSE}), we have the ergodic achievable SE of the $k$th user under the completely overlapped VR case as
\begin{align}\label{Rk_LMMSE_completely_overlap}
R_k^{\text{LMMSE,Com}} &= {\mathbb{E}_{\mathbf{h}}}\left\{ {{{\log }_2}\det \left( {{{\mathbf{I}}_K} + {p_u}{{\mathbf{G}}^H}{\mathbf{\tilde \Theta G}}} \right)} \right\} \nonumber \\&\quad\quad- {\mathbb{E}_{\mathbf{h}}}\left\{ {{{\log }_2}\det \left( {{{\mathbf{I}}_{K - 1}} + {p_u}{\mathbf{G}}_{(k)}^H{\mathbf{\tilde \Theta }}{{\mathbf{G}}_{(k)}}} \right)} \right\} \nonumber \\
        &\mathop  = \limits^{(a)} \frac{{K{{\log }_2}e}}{{\Pi _{m < n}^{\tilde M}({\beta _n} - {\beta _m})}}\bigg( \sum\limits_{i = {\tilde M} - K + 1}^{\tilde M} {\det {{\mathbf{E}}_{K,\tilde M}}(i)} \nonumber \\&\quad\quad - \sum\limits_{i = {\tilde M} - K + 2}^{\tilde M} {\det {{\mathbf{E}}_{K - 1,{\tilde M}}}(i)}  \bigg),
\end{align}
where (a) comes from \cite{Matthew10achievable}, ${\beta _1} >  \ldots  > {\beta _{\tilde M}}$ are the nonzero eigenvalues of ${\mathbf{\tilde \Theta }}$, ${{\mathbf{E}}_{K,\tilde M}}(i)$ and ${{\mathbf{E}}_{K - 1,\tilde M}}(i)$ are $\tilde M\times \tilde M$ matrices with their $(s,t)$th element being
\begin{equation}
{\left[ {{{\mathbf{E}}_{p,\tilde M}}(i)} \right]_{s,t}} = \left\{ {\begin{array}{*{20}{c}}
  {\beta _s^{t - 1},\quad\quad t \ne i}, \\
  {\beta _s^{t - 1}{e^{\frac{1}{{{\beta _s}{p_u}}}}}\sum\nolimits_{h = 1}^{p - \tilde M + t} {{E_h}\left( {\frac{1}{{{\beta _s}{p_u}}}} \right),t = i} },
\end{array}} \right.
\end{equation}
where ${E_h}(\cdot)$ denotes the exponential integral function.

Note that (\ref{Rk_LMMSE_completely_overlap}) is a lower bound of the ergodic achievable SE of the completely overlapped VR case since the multi-user interference is maximized when ${{\mathbf{\Theta }}_1} =  \cdots  = {{\mathbf{\Theta }}_K} = {\mathbf{\Theta }}$.
Besides, due to the existence of VR, only a relatively small block diagonal portion of ${\mathbf{\Theta }}$ will be non-zero. {In this case, the redundant parts are removed according to basic math operations, and the analysis of the non-stationary extra-large massive MIMO system is similar to a stationary massive MIMO system. Hence, the performance analysis of the stationary/traditional massive MIMO can be somewhat regarded as \emph{a special case} of the extra-large scale MIMO when the VRs of different users completely overlap \cite{Carvalho19nonstationary}.}

\emph{\textbf{Partially Overlapped VR Case}}: In this case, users are randomly and relatively sparsely distributed along the whole extra-large scale antenna array and the VRs of different users partially overlap. Theorem\,\ref{Th2} analyzes the ergodic achievable SE for this partially overlapped VR case.\footnote{In fact, when we approximate $({{\mathbf{I}}_K} + {p_u}{{\mathbf{F}}^H}{\mathbf{F}})$ with ${\rm{diag}}({{\mathbf{I}}_K} + {p_u}{{\mathbf{F}}^H}{\mathbf{F}})$, the ergodic achievable SE for the completely overlapped VR case can also be approximated by (\ref{Rk_LMMSE_app_k}) in Theorem\,\ref{Th2} with a looser tightness.}

\begin{Theorem}\label{Th2}
When the LMMSE receiver is employed in the uplink of the  extra-large scale massive MIMO system, the ergodic achievable SE of the $k$th user can be approximated by
\begin{equation}\label{Rk_LMMSE_app_k}
R_k^{\text{LMMSE,app}} = {\log _2}\left[ {1 + {p_u}{\rm{tr}}({\mathbf{B}}{{\mathbf{\Theta }}_k})} \right].
\end{equation}
\end{Theorem}
\begin{IEEEproof}
See Appendix I.
\end{IEEEproof}

Note that in the partially overlapped VR case, there exists a special scenario that only a few users are sparsely distributed in front of the extra-large scale massive MIMO system and, thus, no overlapped VRs appear. The analysis for this special scenario is also presented in Appendix I.
Furthermore, since there is less possibility that VRs of different users completely overlap, especially when cooperated with user scheduling algorithms, we mainly focus on the partially overlapped VR case in the subsequent analysis.

Similar to Theorem\,\ref{Th1}, Theorem\,\ref{Th2} indicates that the ergodic achievable SE per user under LMMSE receiver is also dominated by the matrix product ${\mathbf{B}}{\mathbf{\Theta }}$. Hence, for the purpose of maximizing the system sum achievable SE with the LMMSE receiver, the subarray e.g., the phase coefficients of the phase shifter network, should be well designed to match the correlation matrix ${{\mathbf{\Theta }}}$, such that ${\rm{tr}}({\mathbf{B}}{{\mathbf{\Theta }}})$ is maximized. Based on Theorem\,\ref{Th2}, we also have the following remark.

\begin{Remark}\label{Remark2}
For the LMMSE receiver, in order to maximize the system sum achievable SE, we should simultaneously schedule as many users as possible who have larger ${\rm{tr}}({\mathbf{B}}{{\mathbf{\Theta }}_k})$.
\end{Remark}

Different from the MRC receiver, Remark\,\ref{Remark2} indicates that the LMMSE receiver can effectively cancel the interference from VRs' overlap and thus support more users' communication simultaneously.
In the next section, we precisely elaborate on the phase coefficient design of the subarray with a phase shifter network. Another architecture, i.e., the on-off switch-based subarray, is also investigated to provide insights into the corresponding user scheduling.

\section{Subarray Design}

In this section, we investigate the design of the subarray in the proposed extra-large scale massive MIMO system architecture.
Referring back to the hardware architecture illustrated in Section II, two subarray architectures i.e., the subarray with phase shifters and the on-off switch-based subarray, are considered in this section.

\subsection{Subarray with Phase Shifters}
When subarrays with phase shifters are deployed, every antenna at the BS is connected with an independent phase shifter. In what follows, we consider high precision phase shifters, nevertheless, low resolution phase shifters can also be employed to further reduce the hardware cost which is, however, beyond the scope of this paper. On the basis of (\ref{Rk_MRC_app_k}) in Theorem\,\ref{Th1} and (\ref{Rk_LMMSE_app_k}) in Theorem\,\ref{Th2} and in order to boost the ergodic achievable SE of each user as much as possible, we have the following proposition for the phase coefficient design which seeks to maximize ${\rm{tr}}({\mathbf{B}}{{\mathbf{\Theta }}})$.\footnote{For the MRC receiver in (\ref{Rk_MRC_app_k}), to simplify the analysis, we assume that sophisticated user scheduling algorithms are performed and, thus, the inter-user interference item at the denominator is neglected. Because ${\rm{tr}^2}({\mathbf{B}}{{\mathbf{\Theta }}_k}) > {\rm{tr}}({\mathbf{B}}{{\mathbf{\Theta }}_k}{\mathbf{B}}{{\mathbf{\Theta }}_k})$, we concentrate on maximizing ${\rm{tr}}({\mathbf{B}}{{\mathbf{\Theta }_k}})$ as with the LMMSE receiver in (\ref{Rk_LMMSE_app_k}).}

\begin{Proposition}\label{Pro1}
For subarray with phase shifters, the phase shifter coefficients of the $i$th subarray ${{\mathbf{w}}_i}$ served for user $k$ should be designed as
\begin{equation}\label{w_i}
{{\mathbf{w}}_i} = \frac{N}{M}{e^{j\angle {{\mathbf{v}}_{k,i}}}},
\end{equation}
where $N/M$ is introduced for normalization, ${{\mathbf{v}}_{k,i}}$ is the eigenvector of ${{\mathbf{\bar \Theta }}_{k,ii}}$ corresponding to the maximum eigenvalue, {{$\angle(\cdot)$ represents the angle of its input complex argument}}, and ${{\mathbf{\bar \Theta }}_{k,ii}} \in {\mathbb{C}^{(M/N) \times (M/N)}},\forall i= 1, \ldots ,N$ denotes the $i$th block diagonal matrix of ${{\mathbf{\Theta }}_k}$.
\end{Proposition}
\begin{IEEEproof}
Since ${\mathbf{B}} = \text{blkdiag}({{\mathbf{\bar B}}_1},{{\mathbf{\bar B}}_2}, \ldots ,{{\mathbf{\bar B}}_N})$ and ${{\mathbf{\bar B}}_i} = {{\mathbf{w}}_i}{\mathbf{w}}_i^H,i = 1, \ldots ,N$, we define
\begin{equation}
{{\mathbf{\Theta }}_k} = \left( {\begin{array}{*{20}{c}}
  {{{{\mathbf{\bar \Theta }}}_{k,11}}}& \ldots &{{{{\mathbf{\bar \Theta }}}_{k,1N}}} \\
   \vdots & \ddots & \vdots  \\
  {{{{\mathbf{\bar \Theta }}}_{k,N1}}}& \cdots &{{{{\mathbf{\bar \Theta }}}_{k,NN}}}
\end{array}} \right),
\end{equation}
where ${{\mathbf{\bar \Theta }}_{k,ij}} \in {\mathbb{C}^{(M/N) \times (M/N)}},\forall i,j = 1, \ldots ,N$ denotes the $i$th row $j$th column block matrix of ${{\mathbf{\Theta }}_k}$; then,
\begin{equation}
{\mathbf{B}}{{\mathbf{\Theta }}_k} = \left( {\begin{array}{*{20}{c}}
  {{{{\mathbf{\bar B}}}_1}{{{\mathbf{\bar \Theta }}}_{k,11}}}& \ldots &{{{{\mathbf{\bar B}}}_1}{{{\mathbf{\bar \Theta }}}_{k,1N}}} \\
   \vdots & \ddots & \vdots  \\
  {{{{\mathbf{\bar B}}}_N}{{{\mathbf{\bar \Theta }}}_{k,N1}}}& \cdots &{{{{\mathbf{\bar B}}}_N}{{{\mathbf{\bar \Theta }}}_{k,NN}}}
\end{array}} \right).
\end{equation}
Hence,
\begin{align}\label{tr_phase_shifter}
{\rm{tr}}({\mathbf{B}}{{\mathbf{\Theta }}_k}) &= {\rm{tr}}\left( {\sum\limits_{i = 1}^N {{{{\mathbf{\bar B}}}_i}{{{\mathbf{\bar \Theta }}}_{k,ii}}} } \right) \nonumber \\
                 &\mathop  = \limits^{(a)} \sum\limits_{i \in {S_k}} {{\rm{tr}}({\mathbf{w}}_i^H{{{\mathbf{\bar \Theta }}}_{k,ii}}{{\mathbf{w}}_i})}  \nonumber \\
                 &= \sum\limits_{i \in {S_k}} {{\mathbf{w}}_i^H{{{\mathbf{\bar \Theta }}}_{k,ii}}{{\mathbf{w}}_i}},
\end{align}
where $S_k$ represents the ensemble of the non-zero block matrices ${{\mathbf{\bar \Theta }}_{k,ii}}$ for ${{\mathbf{\Theta }}_{k}}$ and (a) utilizes the trace property ${\rm{tr}}({\mathbf{AB}}) = {\rm{tr}}({\mathbf{BA}})$. Consequently, based on (\ref{tr_phase_shifter}), to maximize the ergodic achievable SE for user $k$, ${{\mathbf{w}}_i}$ should be chosen as the eigenvector of ${{\mathbf{\bar \Theta }}_{k,ii}}$ corresponding to the maximum eigenvalue. Similar to \cite{Payami16Hybrid,Payami19Phase}, considering that ${{\mathbf{w}}_i}$ is realized by phase shifters with constant-modulus constraints, we design ${{\mathbf{w}}_i}$ as in (\ref{w_i}).
\end{IEEEproof}

It is important to note that the phase coefficient design of each subarray in (\ref{w_i}) is designed in terms of the low-dimension matrix ${{\mathbf{\bar \Theta }}_{k,ii}}\in {\mathbb{C}^{(M/N) \times (M/N)}}$, instead of the correlation matrix ${{\mathbf{\Theta }}_{k}}\in {\mathbb{C}^{M \times M}}$ across the entire extra-large scale antenna array. Therefore, the calculation of ${{\mathbf{w}}_i}$ for different subarrays can be executed in parallel and the computation complexity can also be greatly reduced.

\subsection{On-Off Switch-Based Subarray}
At the expense of performance degradation, an on-off switch-based subarray requires much lower hardware complexity and hardware cost when compared with a subarray with phase shifters. {To investigate how to maximize the ergodic achievable SE when on-off switch-based subarrays are configured, we first assume that the switches are all turned on in the uplink, and} ${\mathbf{B}}$ for the on-off switch-based subarray becomes ${\mathbf{B}} = \frac{N}{M}{\rm{diag}}({{\mathbf{1}}_{M/N}},{{\mathbf{1}}_{M/N}}, \ldots ,{{\mathbf{1}}_{M/N}})$,
where ${{\mathbf{1}}_{M/N}}$ denotes the all-ones matrix.
Suppose ${\mathbf{\Lambda }} = {\rm{blkdiag}}({{\mathbf{1}}_{M/N}},{{\mathbf{1}}_{M/N}}, \ldots ,{{\mathbf{1}}_{M/N}})$, then ${\mathbf{B}} = {N{\mathbf{\Lambda }}}/{M}$, such that (\ref{Rk_MRC_app_k}) and (\ref{Rk_LMMSE_app_k}) can be simplified to
\begin{equation}\label{eq:RkMRCappkonoff}
R_k^{\text{MRC,app}} = {\log _2}\left( {1 + \frac{{{\rm{tr}}({\mathbf{\Lambda }}{{\mathbf{\Theta }}_k}{\mathbf{\Lambda }}{{\mathbf{\Theta }}_k}) + {\rm{t}}{{\rm{r}}^2}({\mathbf{\Lambda }}{{\mathbf{\Theta }}_k})}}{{\sum\limits_{i = 1,i \ne k}^K {{\rm{tr}}({\mathbf{\Lambda }}{{\mathbf{\Theta }}_i}{\mathbf{\Lambda }}{{\mathbf{\Theta }}_k})}  + \frac{M}{{{p_u}N}}{\rm{tr}}({\mathbf{\Lambda }}{{\mathbf{\Theta }}_k})}}} \right)
\end{equation}
and
\begin{equation}
R_k^{\text{LMMSE,app}} = {\log _2}\left[ {1 + \frac{N{p_u}{\rm{tr}}({\mathbf{\Lambda }}{{\mathbf{\Theta }}_k})}{M}} \right]
\end{equation}
respectively. As with the case of subarray with phase shifters, to maximize the ergodic achievable SE, we pay attention to the analysis of ${\rm{tr}}({\mathbf{\Lambda }}{{\mathbf{\Theta }}_k})$ as well. In the on-off switch-based subarray, we have
\begin{align}\label{eq:onoffsubarrayTrace}
{\text{tr}}({\mathbf{\Lambda }}{{\mathbf{\Theta }}_k}) 
                 &= \sum\limits_{i \in {S_k}} {\sum\limits_{m < n} {2\operatorname{Re} \left( {{{\left( {{{{\mathbf{\bar \Theta }}}_{k,ii}}} \right)}_{mn}}} \right)} }  + {\rm{tr}}({{\mathbf{\Theta }}_k}),
\end{align}
where ${\left( {{{{\mathbf{\bar \Theta }}}_{k,ii}}} \right)_{mn}}$ denotes the $m$th row $n$th column element of ${{\mathbf{\bar \Theta }}_{k,ii}}$ and ${\operatorname{Re} \left( {{{\left( {{{{\mathbf{\bar \Theta }}}_{k,ii}}} \right)}_{mn}}} \right)}$ represents the real part of ${\left( {{{{\mathbf{\bar \Theta }}}_{k,ii}}} \right)_{mn}}$. Hence, $\sum\limits_{i \in {S_k}} {\sum\limits_{m < n} {2\operatorname{Re} \left( {{{\left( {{{{\mathbf{\bar \Theta }}}_{k,ii}}} \right)}_{mn}}} \right)}} + {\rm{tr}}({{\mathbf{\Theta }}_k})$ is the sum of the real parts of the $M/N$-dimension non-zero main block diagonal matrices of the $k$th user's correlation matrix. This indicates that ${\rm{tr}}({\mathbf{\Lambda }}{{\mathbf{\Theta }}_k})$, to some extent, reflects the sum power that has been radiated on the on-off switch-based subarrays at the BS by the $k$th user. As a consequence, we obtain the following proposition for user scheduling when on-off switch-based subarrays are configured.
\begin{Proposition}\label{Pro2}
When on-off switch-based subarrays are deployed at the extra-large scale massive MIMO system, the user who has larger sum energy radiated to the subarrays should be scheduled for communication in order to maximize the system sum achievable SE.
\end{Proposition}

{Moreover, combining (\ref{eq:onoffsubarrayTrace}) and the existence of VR, we observe that, for the on-off switch-based subarray architecture, only the switches in the energy-dominant subarrays are needed to be turned on while the others can be turned off during the uplink data transmission. This conclusion will be leveraged in the proposed user scheduling algorithm and be manifested in Section VI.C.}

\section{User Scheduling}
Based on the acquired ergodic achievable SE approximations provided in Theorems\,\ref{Th1} and \ref{Th2} as well as the subarray design provided in Propositions\,\ref{Pro1} and \ref{Pro2}, we propose two user scheduling algorithms for the extra-large scale massive MIMO system in this section.

Scheduling is of great significance in multi-user communication systems, especially for extra-large scale massive MIMO because of the existence of VRs, which could be used for further improving the spectral and energy efficiency.
However, instantaneous CSI-based scheduling is practically challenging for extra-large scale massive MIMO due to the unconventionally large number of antenna elements and relatively large number of users to be served.
To tackle this problem, on the basis of Theorems\,\ref{Th1} and \ref{Th2}, we propose two statistical CSI-based greedy scheduling schemes with the aim of maximizing the system sum achievable SE. Given that the MRC receiver is adopted, Remark\,\ref{Remark1} indicates that, users whose VRs cover different subarrays or those with fewer overlapped VRs should be scheduled so as to maximize the sum achievable SE. In the other case, where a LMMSE receiver is employed, Remark\,\ref{Remark2} showcases that, to obtain the maximum of the achievable SE, as many users as possible with larger ${\rm{tr}}({\bf{B}}{{\bf{\Theta }}_i})$ should be scheduled. In the following, we firstly schedule users utilizing statistical CSI in a greedy manner. Then, the algorithm investigating the feasibility of jointly scheduling users and subarrays after taking energy consumption into consideration is provided.

\subsection{Statistical CSI-based Greedy User Scheduling}
Generally, utilizing exhaustive search could return the optimal solution in scheduling problems, however, it may not be appropriate for  extra-large scale massive MIMO due to the extremely large computational complexity and long runtime. Therefore, a sub-optimal user scheduling algorithm, i.e., the greedy user scheduling algorithm, whose main idea is to achieve an optimal result during each scheduling step and, thus, greatly reduce the algorithm complexity, is employed. We summarize our proposed statistical CSI-based greedy user scheduling algorithm in {\bf Algorithm\,\ref{Alogrithm1}}.

\begin{algorithm}[h]
\small
	\caption{Statistical CSI-based Greedy User Scheduling Algorithm}\label{Alogrithm1}
	\begin{algorithmic}[1]
\REQUIRE  ${U_s} = \emptyset$, ${U_n} = \{ 1,2, \ldots ,K\}$, ${N_s} = 0$, ${N_u}$, $R = 0$, $R_{temp} =0$. \\	
\WHILE {${N_s} < {N_u}$}
		\FOR{each ${u_i} \in {U_n}$}
		\STATE calculate the system sum achievable SE ${R_{U_s\cup{u_i}}}$;\\		
		\ENDFOR
		\STATE select ${u_{i}}$ with the largest $R$ among ${R_{U_s\cup{u_i}}}$ as a newly scheduled user candidate ${u_{sel}}$;
		\IF {$R_{temp} \leq R$}
		\STATE ${U_s} = {U_s} \cup {u_{sel}}, {U_n} = {U_n}\backslash \{ {u_{sel}}\}, R_{temp}=R, {N_s} = {N_s} + 1$;\\
		\ELSE
		\STATE {break};\\
		\ENDIF
	    \STATE {$R=R_{temp}$};\\
		\ENDWHILE
		\ENSURE ${U_s}, R$.
	\end{algorithmic}
\end{algorithm}

In {\bf Algorithm\,\ref{Alogrithm1}}, firstly, we initialize all the system parameters, including the scheduled user set ${U_s} = \emptyset$, the number of scheduled users ${N_s} = 0$, and the unscheduled user set ${U_n} = \{ 1,2, \ldots ,K\}$. The total number of users to be scheduled and served is ${N_u}$, and the system sum achievable SE is initialized as $R=R_{temp} = 0$.

Next, we select the users in the unscheduled user set ${U_n}$ one by one and calculate their corresponding updated system sum achievable SEs based on the scheduling results of the previous iteration. A user will be added to the scheduled user set only if it reaches the maximum of the updated system sum achievable SEs among all unscheduled users from ${U_n}$, as well as produces a positive gain compared with the last iteration results. To update the system sum achievable SE in this step, (\ref{Rk_MRC_app_k}) and (\ref{Rk_LMMSE_app_k}) are leveraged when MRC and LMMSE receivers are respectively considered.
Note that in the phase shifter-based subarray, if a subarray is not covered by any user's VR, the phase coefficients of the subarray will be set to the default value zero, i.e., $\angle {{\mathbf{w}}_{i}}=\mathbf{0}$; if a subarray is covered by multiple users simultaneously, then the phase coefficients of the subarray will be set to the sum of the phases corresponding to the multiple users.

Then, the algorithm keeps running until ${N_s} = {N_u}$ or there is no SE gain when adding a new user. Finally, the algorithm outputs the final scheduling results i.e., the scheduled user set ${U_s}$ and the system sum achievable SE $R$.

Note that the proposed greedy user scheduling algorithm exploits only statistical CSI, i.e., the knowledge of the channel correlation information instead of the instantaneous channel gains. This is beneficial and more practical especially for extra-large scale massive MIMO systems. Moreover, considering the existence of VR and the cases that some subarrays may be covered by no user, we further examine the possibility to jointly schedule users and subarrays and propose a statistical CSI-based greedy joint user and subarray scheduling scheme in the next subsection.

\subsection{Statistical CSI-based Greedy Joint User and Subarray Scheduling}
Different from {\bf Algorithm\,\ref{Alogrithm1}}, subarrays are jointly scheduled in the proposed statistical CSI-based greedy joint user and subarray scheduling algorithm i.e., {\bf Algorithm\,\ref{Alogrithm2}}.
Taking into consideration that each user only covers a limited portion of antenna arrays of BS and that some BS subarrays are possibly covered by no user, {\bf Algorithm\,\ref{Alogrithm2}} can significantly enhance the energy efficiency by turning off uncovered subarrays, thereby facilitating the practical implementation of extra-large scale massive MIMO systems. The details of {\bf Algorithm\,\ref{Alogrithm2}} are presented in the following.

\begin{algorithm}[h]
\small
	\caption{Statistical CSI-based Greedy Joint User and Subarray Scheduling Algorithm}\label{Alogrithm2}
	\begin{algorithmic}[1]
		\REQUIRE
		 ${U_s} = \emptyset$, $S = \emptyset$, ${U_n} = \{ 1,2, \ldots ,K\}$, ${S_n} = \{ 1,2, \ldots ,N\}$, ${N_s} = 0$, ${N_u}$, $R = 0$, ${Sub_{\max }}$, ${Sub_{\min }}$, $R_{temp} =0$. \\	
		\WHILE {${N_s} < {N_u}$}
		\FOR{each ${u_i} \in {U_n}$}
			\FOR{each ${S_{u_i}} \subset  {S_n}$}
				\IF{$Su{b_{\min }} \le {|S_{u_i}|} \le  Su{b_{\max }}$ }
		\STATE calculate the system sum achievable SE ${R_{U_s\cup{u_i},S_{u_i}}}$;\\
		\ELSE
		\STATE {continue};
		\ENDIF
		\ENDFOR	
		\STATE {select ${S_{sel,u_i}}$ with the largest $R$ among ${R_{U_s\cup{u_i},S_{u_i}}}$  as ${u_i}$'s subarray candidate;}
		\ENDFOR
		\STATE select ${u_{sel}}$ with the largest $R$ among ${R_{U_s\cup{u_i},S_{sel,u_i}}}$  as a newly scheduled user candidate;\\
		\IF {$R_{temp} \leq {R_{U_s\cup{u_{sel}},S_{sel,u_{sel}}}}$}
		\STATE ${U_s} = {U_s} \cup {u_{sel}}, S = S \cup {S_{sel,u_{sel}}}, {U_n} = {U_n}\backslash \{ {u_{sel}}\}, {S_n} = {S_n}\backslash {S_{sel,u_{sel}}},  R_{temp}=R$, {${N_s} = {N_s} + 1$};\\
		\label{code:TrainBase:pos}
		\ELSE
		\STATE {break};
		\ENDIF
		\STATE {$R=R_{temp}$};
		\ENDWHILE
		\ENSURE ${U_s}, S, R$.
	\end{algorithmic}
\end{algorithm}

We initialize all the system parameters at the first step in {\bf Algorithm\,\ref{Alogrithm2}}, including the scheduled user set ${U_s} = \emptyset$, the scheduled subarray set $S = \emptyset$, the unscheduled user set ${U_n} = \{ 1,2, \ldots ,K\}$, the unscheduled subarray set ${S_n} = \{ 1,2, \ldots ,N\}$, the number of scheduled users ${N_s} = 0$, the total number of users to be scheduled and served ${N_u}$, and the system sum achievable SE $R=R_{temp} = 0$.
The maximum and minimum number of scheduled subarrays per user are set as ${Sub_{\max }}$ and ${Sub_{\min }}$, respectively.

Next, for each user in the unscheduled user set ${U_n}$, we select its best subarray set from the unscheduled subarray set ${S_n}$ (the number of selected subarrays must not be greater than ${Sub_{\max }}$ and less than ${Sub_{\min }}$) for transmission and receive combining, so that the system sum achievable SE is maximized after the current user is added.
If the updated system sum achievable SE is larger than its counterpart, then we record the corresponding user index with its selected subarray set and the updated system sum achievable SE. As a result, this user becomes a candidate.

Based on the obtained candidates, the user who contributes with the strongest gain to the sum achievable SE is finally selected and added to ${U_s}$, with its corresponding selected subarray set $S_{sel,u_{sel}}$ added to $S$. At the same time, its user index $u_{sel}$ and selected subarray set $S_{sel,u_{sel}}$ are removed from $U_n$ and $S_n$ respectively. After that, the number of scheduled users, i.e., ${N_s}$, increases by one. When ${N_s} = {N_u}$ or there is no SE gain when adding a new user, the scheduling algorithm terminates and outputs ${U_s}$, $S$ and $R$. The next section demonstrates the performance of the proposed greedy joint user and subarray scheduling algorithm.

\section{Numerical Results}
In this section, the tightness of the approximated uplink ergodic achievable SEs under both MRC and LMMSE receivers is firstly investigated. Then, we verify the effectiveness of the proposed phase coefficient design in Section IV. {After that, we demonstrate the energy efficiency of different architectures.}
The performance of these proposed two statistical CSI-based user scheduling algorithms, i.e., the statistical CSI-based greedy user scheduling algorithm and the statistical CSI-based greedy joint user and subarray scheduling algorithm, is also evaluated.
During the simulation, the channel model presented in Section II.B is adopted, with the antenna element spacing being half carrier wavelength and the non-zero diagonal elements of ${{\bf{D}}_k}$ being 1.

\begin{table}[h]
\caption{Values of Main Parameters Used in Numerical Simulations.}\label{tab:parameters}
\small
\centering
  \begin{tabular}{ccccccc}
    \toprule[1.2pt]
    {Parameter} & {Fig.\,2}& {Fig.\,3}& {Fig.\,4}& {Fig.\,5} & {Fig.\,6} & {Fig.\,10} \\
    \hline
    $M$ & $1024$& $1024$& $1024$& $1024$& $1024$ & $1024$\\
    \rowcolor{mygray}
    $N$ & $128$ & $128$ & $128$ & $128$ &$128$   & $128$\\
    $K$ & $20$  & $10$  & $5$   & $5$   & $22$   & $22$\\
    \rowcolor{mygray}
    $E$ & $160$ & $128$ & $160$ & $160$ &$128$   & $128$\\
    \bottomrule[1.0pt]
    \hline
  \end{tabular}
\end{table}
\vspace{-0.5cm}
\subsection{Tightness of the Approximated Uplink Ergodic Achievable SE}
\begin{figure}[h]
\centering
\includegraphics[width=3.4in,height=2.7in]{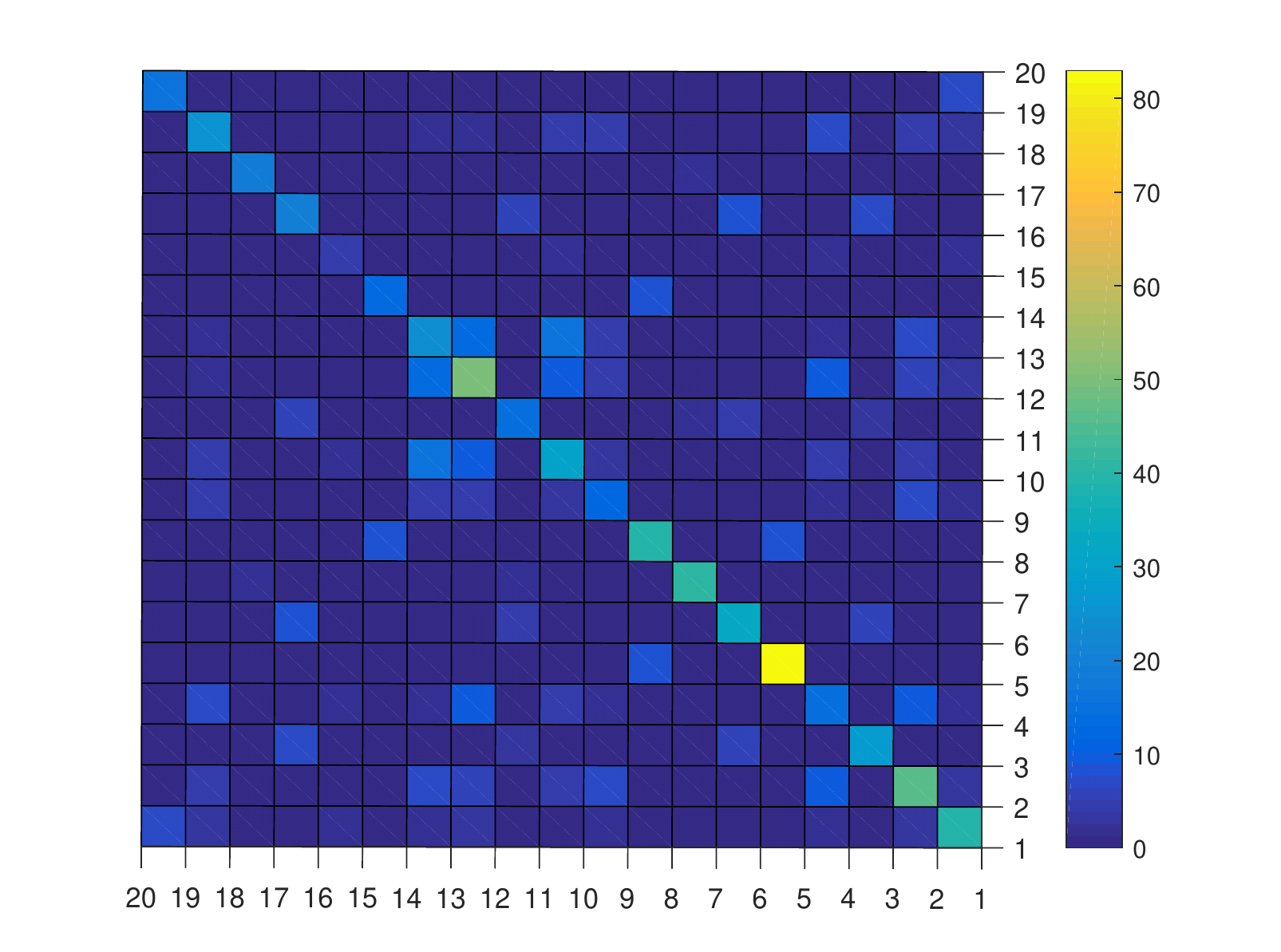}
\caption{ The amplitudes of all the elements in ${\mathbf{Z}}$ when $K=20$ and all the users are randomly located along the extra-large antenna array with each user's VR covering $160$ antenna elements, i.e., $E=160$.}\label{fig:1}
\end{figure}

{Since the diagonal-dominant property of ${\mathbf{Z}}$ plays an important role in the derivation of the approximated uplink achievable SEs, we first verify it in Fig.\,\ref{fig:1}.} Fig.\,\ref{fig:1} provides the amplitudes of all the elements in ${\mathbf{Z}}$ when $K=20$ where all the users are randomly located along the extra-large antenna array with each user's VR covering $160$ antenna elements, i.e., $E=160$.\footnote{For simplicity, we assume that each user's VR covers the same number of antenna elements. However, the simulation methodology also supports the general case, i.e., different $E$ for different users.} Table\,\ref{tab:parameters} summarizes the values of the main parameters used in the numerical simulations for each figure.
As can be seen from Fig.\,\ref{fig:1}, the diagonal elements of ${\mathbf{Z}}$ are apparently larger than the off-diagonal ones, which verifies the conclusion we drew in Section III.B.

\begin{figure}[h]
\centering
\includegraphics[width=3.6in,height=2.9in]{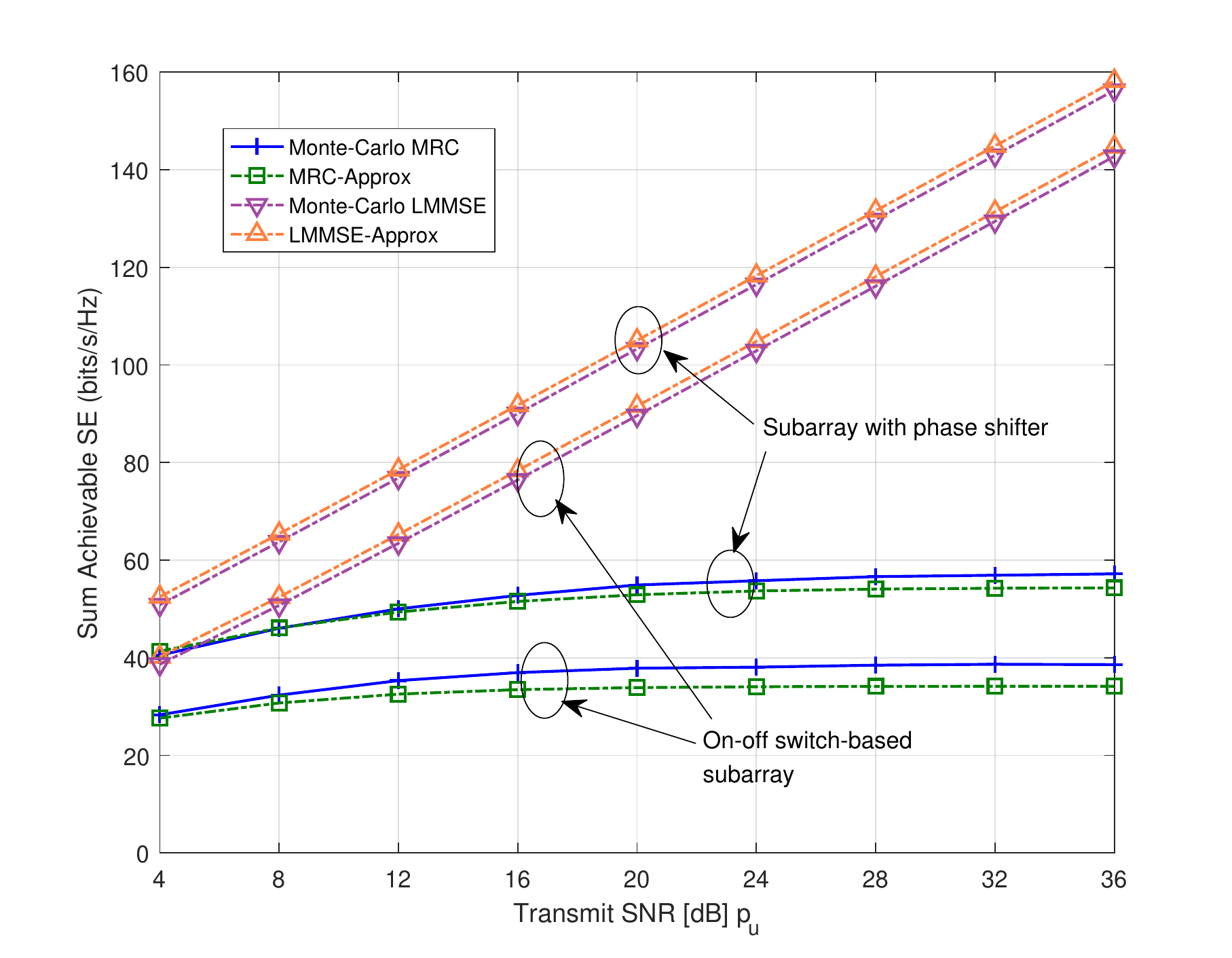}
\caption{The uplink sum achievable SEs with phase shifter-based subarrays and on-off switch-based subarrays: $M=1024$, $N=128$, $K=10$, and $E=128$, and users are randomly located along the extra-large antenna array without user scheduling.}\label{fig:2}
\end{figure}

Next, we investigate the tightness of the approximated uplink achievable SEs. Fig.\,\ref{fig:2} presents the uplink sum achievable SEs under the architectures of the phase shifter-based subarray and the on-off switch-based subarray. Users are randomly located along the extra-large antenna array without user scheduling.
Note that in the phase shifter-based subarray, if a subarray is not covered by any user's VR, the phase coefficients of the subarray would be set to zero by default; otherwise, if a subarray is covered by multiple users simultaneously, then the phase coefficients of the subarray would be set to the sum of the phases corresponding to the multiple users. {For the on-off switch-based subarray, and unless otherwise discussed, we assume all the switches are turned on in the simulation for simplicity.}

From Fig.\,\ref{fig:2}, the proposed achievable SE approximations match well with the Monte-Carlo results.\footnote{The Monte-Carlo results are obtained according to (\ref{Rk_general}).} This indicates that the proposed achievable SE approximations in (\ref{Rk_MRC_app_k}) and (\ref{Rk_LMMSE_app_k}) are inherently useful for the subsequent user scheduling to maximize the system sum achievable SE. In addition, as the power of transmitted signal increases, the system sum achievable SEs continuously increase with the LMMSE receiver since it can effectively eliminate the interference between different users. However, for the MRC receiver, the system sum achievable SEs rapidly tend to saturation due to the persistent inter-user interference.

Furthermore, comparing these results in Fig.\,\ref{fig:2}, we find that, although the on-off switch-based subarray has an apparent performance loss in comparison to the structure of subarray with phase shifters, it can still achieve nearly $70\%$ spectral efficiency performance with the MRC receiver and $80\%$ with the LMMSE receiver. For example, with the MRC receiver, the saturated sum achievable SE under the phase shifter-based subarray architecture is $60$\,bits/s/Hz, while the saturated sum achievable SE with the on-off switch-based subarray is about $40$\,bits/s/Hz; for the LMMSE receiver at the transmit SNR of $36$\,dB, the system sum achievable SEs are $142.7$\,bits/s/Hz and $156.3$\,bits/s/Hz under the on-off switch-based subarray and phase-shifter-based subarray, respectively.
Moreover, the on-off switch-based subarray architecture can effectively reduce the hardware cost of the system by replacing the expensive phase shifters with low-cost switches.
Hence, it would be a more practical hardware solution to apply the on-off switch-based subarray architecture in extra-large scale massive MIMO.

\begin{figure}[h]
\centering
\includegraphics[width=3.6in,height=2.9in]{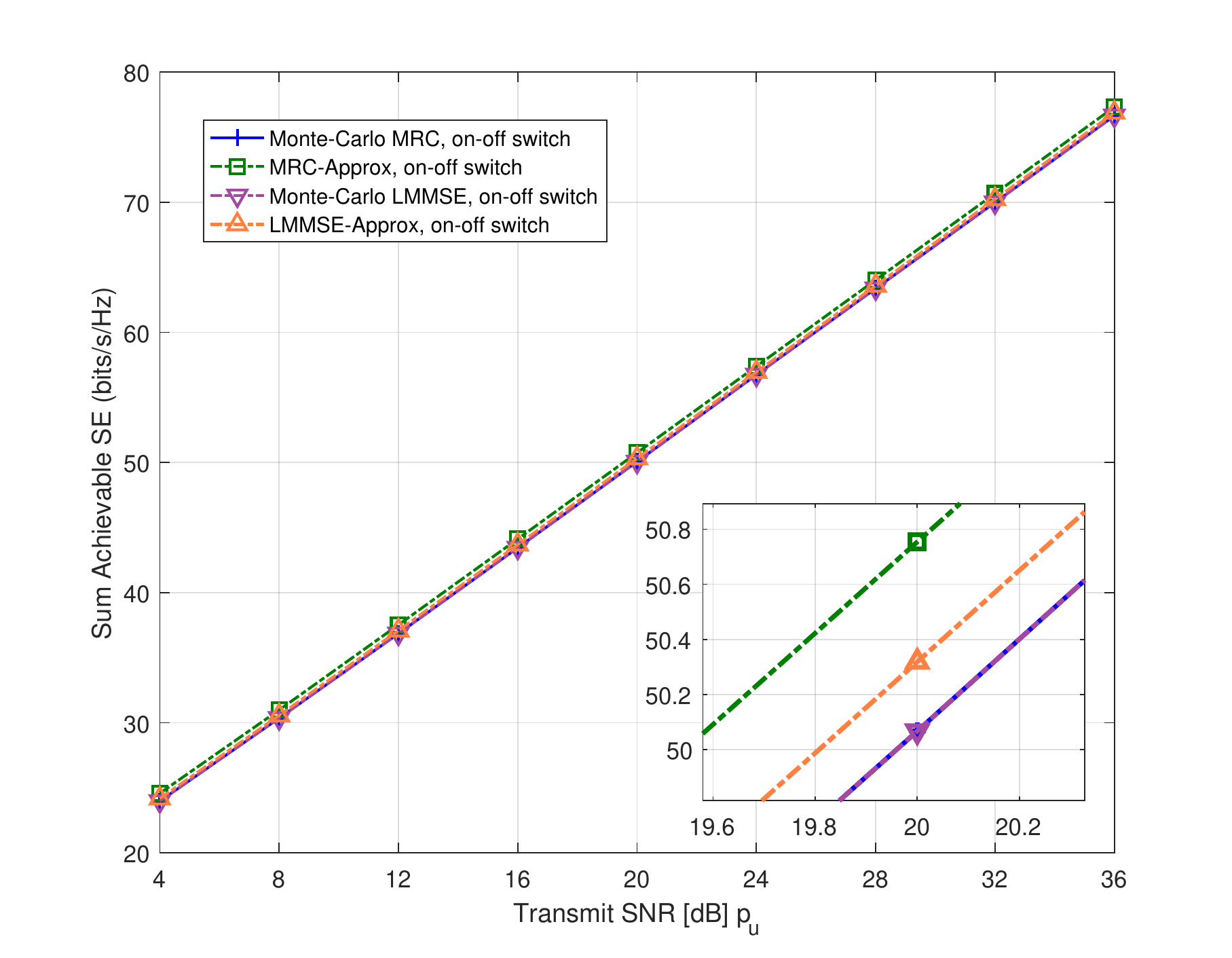}
\caption{ The uplink sum achievable SEs under the scenario of no VR overlapping: $M=1024$, $N=128$, $K=5$, and $E=160$. The architecture of the on-off switch-based subarray is considered.}\label{fig:4}
\end{figure}

Fig.\,\ref{fig:4} presents the sum achievable SE under the special scenario of no overlapped VR, namely users are far apart from each other and the signal radiated by a different user covers different portions of the antenna array. The on-off switch-based subarray architecture is considered.
As can be observed in Fig.\,\ref{fig:4}, the proposed achievable SE approximations (\ref{Rk_MRC_app_k}) and (\ref{Rk_LMMSE_app_k}) yield again great tightness. Moreover, since users' VRs do not overlap and, thus, no interference exists between users, the system sum achievable SE under the MRC receiver continuously increases with an increasing transmit power. Consequently, the MRC receiver achieves the same spectral efficiency as the LMMSE receiver.
Hence, when there are fewer users to be served or when the scheduled users have no overlapped VR, the hardware-friendly MRC receiver should be considered, thereby achieving lower computational complexity with satisfactory performance.

\begin{figure}[h]
\centering
\includegraphics[width=3.6in,height=2.9in]{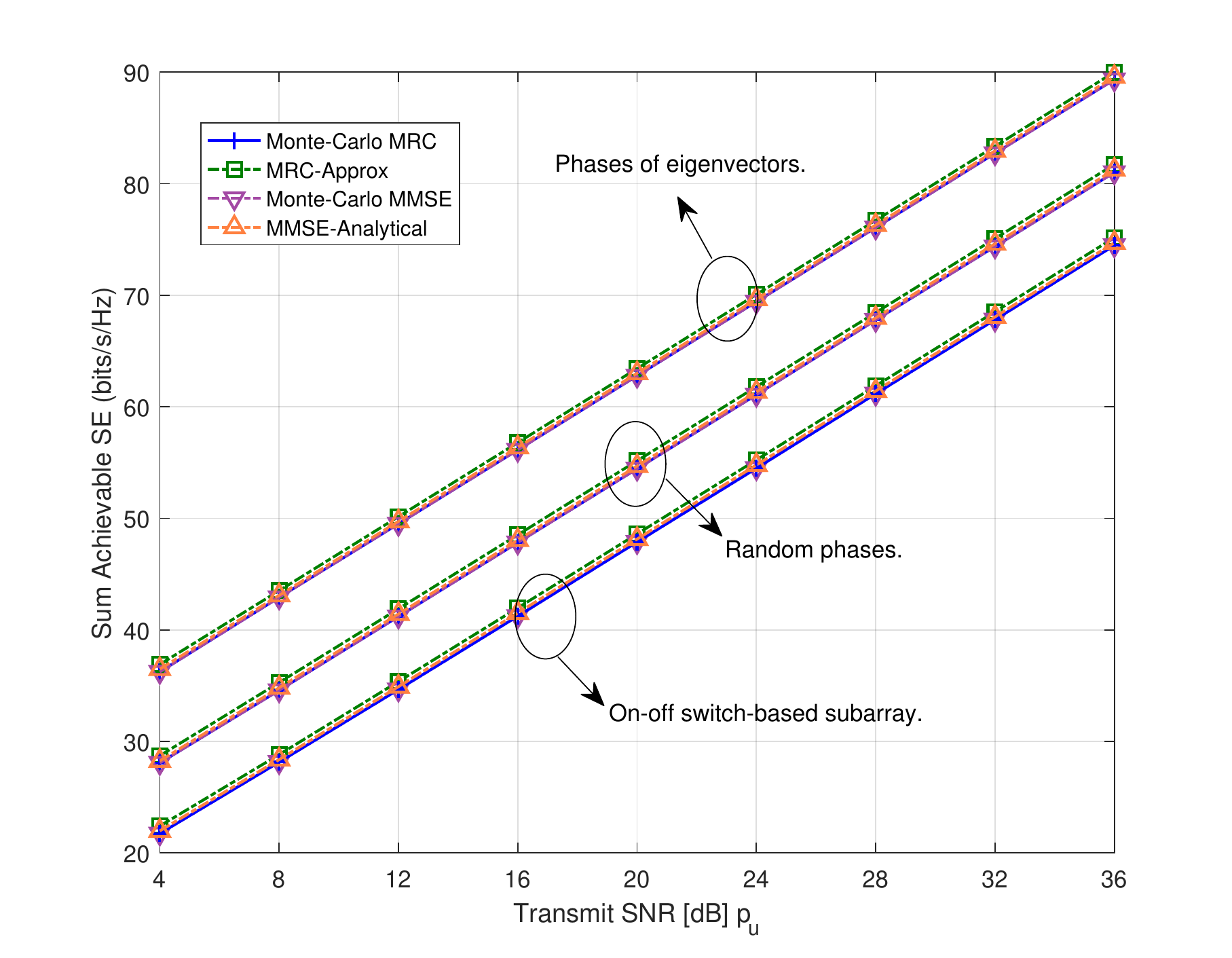}
\caption{ Comparison of the different subarray phase coefficient design under the scenario of no VR overlapping: $M=1024$, $N=128$, $K=5$, and $E=160$.}\label{fig:5}
\end{figure}

\subsection{Comparison of Different Subarray Phase Coefficient Design}
The sum achievable SEs under the special scenario of no overlapped VR for both the phase shifter-based subarray architecture and the on-off switch-based subarray architecture are provided in Fig.\,\ref{fig:5}. The LMMSE receiver is employed and two phase coefficient designs i.e., the proposed eigenvector-based phase coefficient design in Section IV and the random phase coefficient design, are considered for the phase shifter-based subarray.
The results in Fig.\,\ref{fig:5} indicate that the proposed eigenvector-based phase coefficient design achieves the best performance and reaps about $9$\,bits/s/Hz and $14$\,bits/s/Hz sum achievable SE gains over the random phase coefficient design and the on-off switch-based subarray design, respectively. What is more, due to the lack of phase alignment, the on-off switch-based subarray has the lowest system sum achievable SE. However, the on-off switch-based subarray induces the lowest hardware cost and computation complexity, therefore offering a low-cost alternative.
Additionally, even if we adopt the on-off switch-based subarray architecture, $48$\,bits/s/Hz system sum achievable SEs can still be achieved at the SNR of $20$\,dB, which means that the averaged achievable SEs per user are $9.6$\,bits/s/Hz.

\subsection{Energy Efficiency Analysis}
\begin{figure}[h]
\centering
\includegraphics[width=3.6in,height=2.9in]{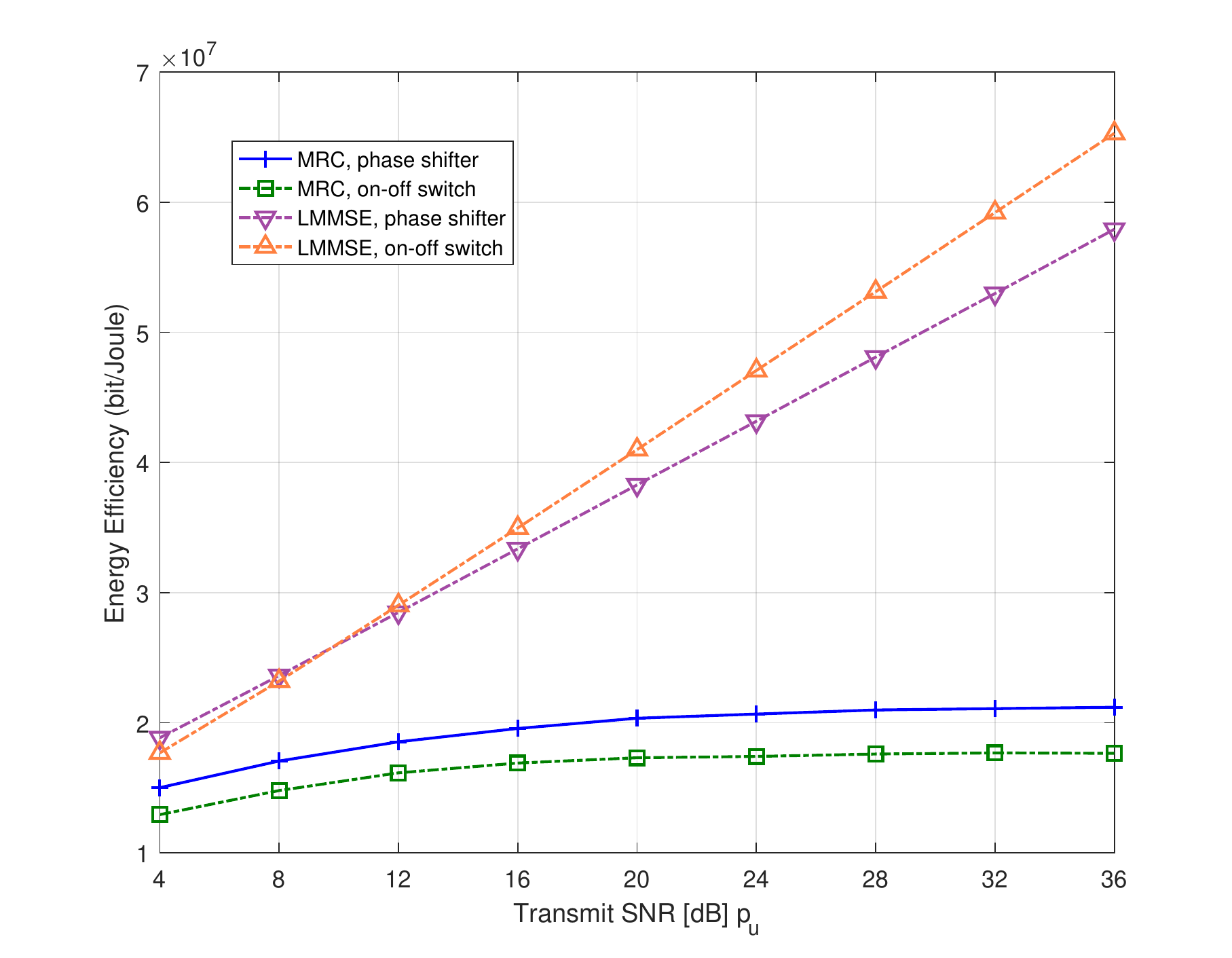}
\caption{Energy efficiency comparison of the different subarray architectures. $M=1024$, $N=128$ and $B=20$\,MHz.}\label{fig:14}
\end{figure}

In this subsection, we demonstrate the energy efficiency analysis for different subarray architectures.
Similar to \cite{Rial15Channel}, we can construct the energy consumption model for different architectures as follows:
\begin{equation}\label{eq:PUPhasesys}
P_{\text{P}}^{\text{U}} = M{P_{{\text{Phase shifter}}}}+N\left( {{P_{{\text{LNA}}}} + {P_{{\text{RF chain}}}} + {P_{{\text{ADC}}}}} \right) + P_{{\text{BB}}}^{\text{U}},\nonumber
\end{equation}
\begin{equation}\label{eq:PUSwitchsys}
P_{\text{S}}^{\text{U}} = M{P_{{\text{Switch}}}}+N\left( {{P_{{\text{LNA}}}} + {P_{{\text{RF chain}}}} + {P_{{\text{ADC}}}}} \right) + P_{{\text{BB}}}^{\text{U}},\nonumber
\end{equation}
where $P_{\text{P}}^{\text{U}}$ and $P_{\text{S}}^{\text{U}}$ denote the uplink system power consumption of the phase shifter-based architecture and the on-off switch-based architecture, respectively; ${P_{{\text{Phase shifter}}}}$ and $P_{\text{Switch}}$ represent the power consumption of phase shifter and switch, respectively; ${P_{{\text{LNA}}}}$ is the low noise amplifier's power consumption; ${P_{{\text{RF chain}}}}$ denotes the power consumption of the RF components, including mixer, filter, and so on; ${P_{{\text{ADC}}}}$ represents the ADC's power consumption, and $P_{{\text{BB}}}^{\text{U}}$ is the power consumption for uplink baseband signal processing. Similarly, we can also set
\begin{align}
{P_{{\text{Phase shifter}}}} &= 20\,{\text{mW}},\quad
{P_{{\text{Switch}}}} = 10\,{\text{mW}},\nonumber\\
P_{\text{LNA}}&=20\,{\text{mW}},\quad
{P_{{\text{RF chain}}}} = 40\,{\text{mW}},\nonumber\\
{P_{{\text{ADC}}}} &= 200\,{\text{mW}},\quad
P_{{\text{BB}}}^{\text{U}} = 200\,{\text{mW}}.\nonumber
\end{align}
Then, the system uplink energy efficiencies of the phase shifter-based subarray architecture and the on-off switch-based subarray architecture can be calculated as
\begin{align}
\eta _{{\text{EE}}}^{\text{P}}& = B\frac{{R _{{\text{MRC/LMMSE}}}^{\text{P}}}}{{P_{\text{P}}^{\text{U}}}}{\text{ (bits/Joule)}},\nonumber\\
\eta _{{\text{EE}}}^{\text{S}} &= B\frac{{R _{{\text{MRC/LMMSE}}}^{\text{S}}}}{{P_{\text{S}}^{\text{U}}}}{\text{ (bits/Joule)}},\nonumber
\end{align}
respectively, where $B$ denotes the system transmission bandwidth.
Fig.\,\ref{fig:14} presents the energy efficiency of different architectures with simulation parameters consistent with Fig.\,\ref{fig:2}.

As can be observed in Fig.\,\ref{fig:14}, the on-off switch-based architecture can achieve better energy efficiency than the phase shifter-based architecture when a LMMSE receiver is utilized. This further manifests the advantages of the former.

\subsection{User Scheduling}
\begin{figure}[h]
\centering
\includegraphics[width=3.6in,height=2.9in]{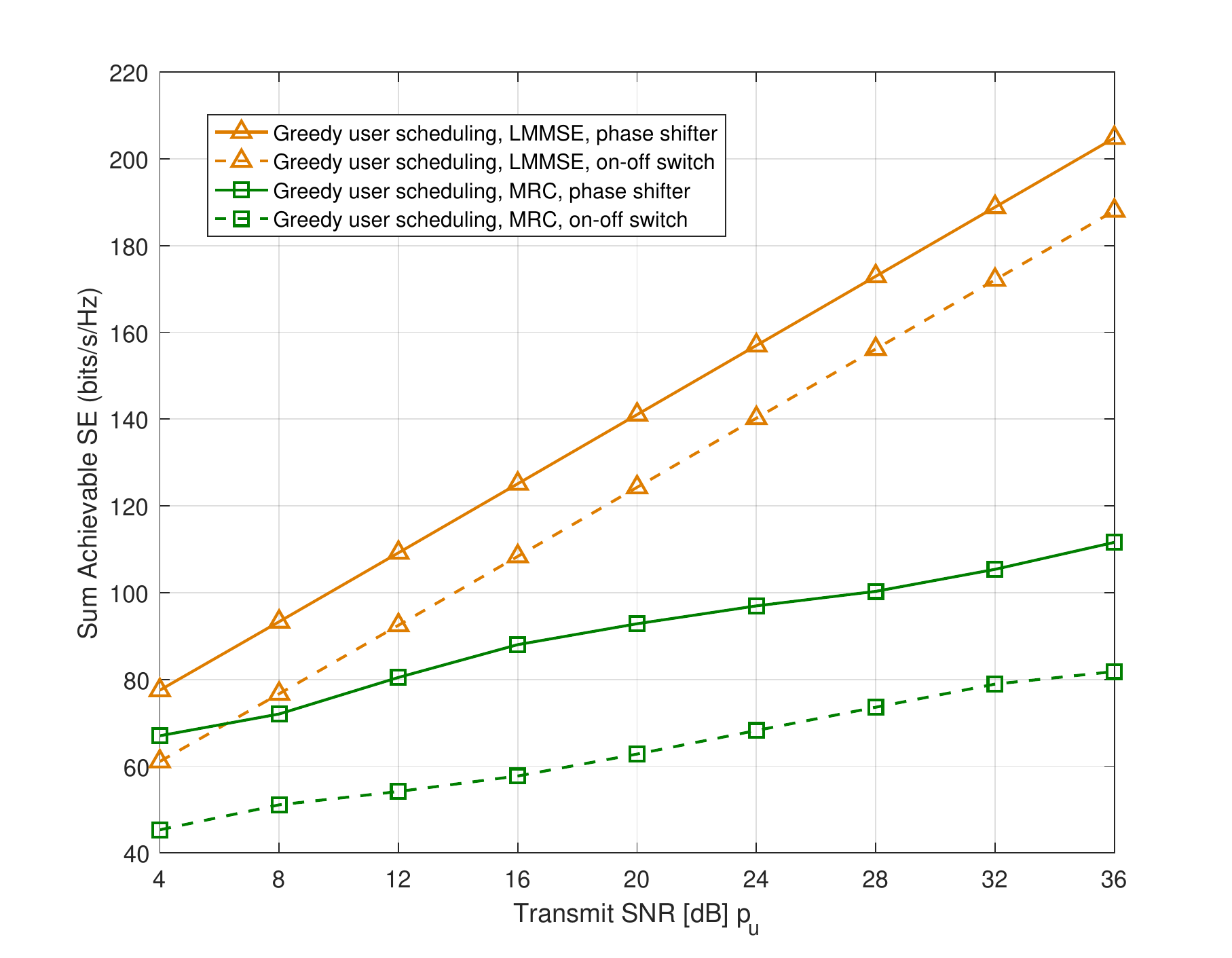}
\caption{ The uplink sum achievable SEs with a phase shifter-based subarray and the on-off switch-based subarray. The parameters $M=1024$, $N=128$, $K=22$, and $E=128$ are selected and users are randomly located along the extra-large antenna array. The statistical CSI-based greedy user scheduling algorithm is utilized and the number of users to be scheduled and served is $N_u=12$.}\label{fig:6}
\end{figure}

\begin{figure}[h]
\centering
\includegraphics[width=3.6in,height=2.9in]{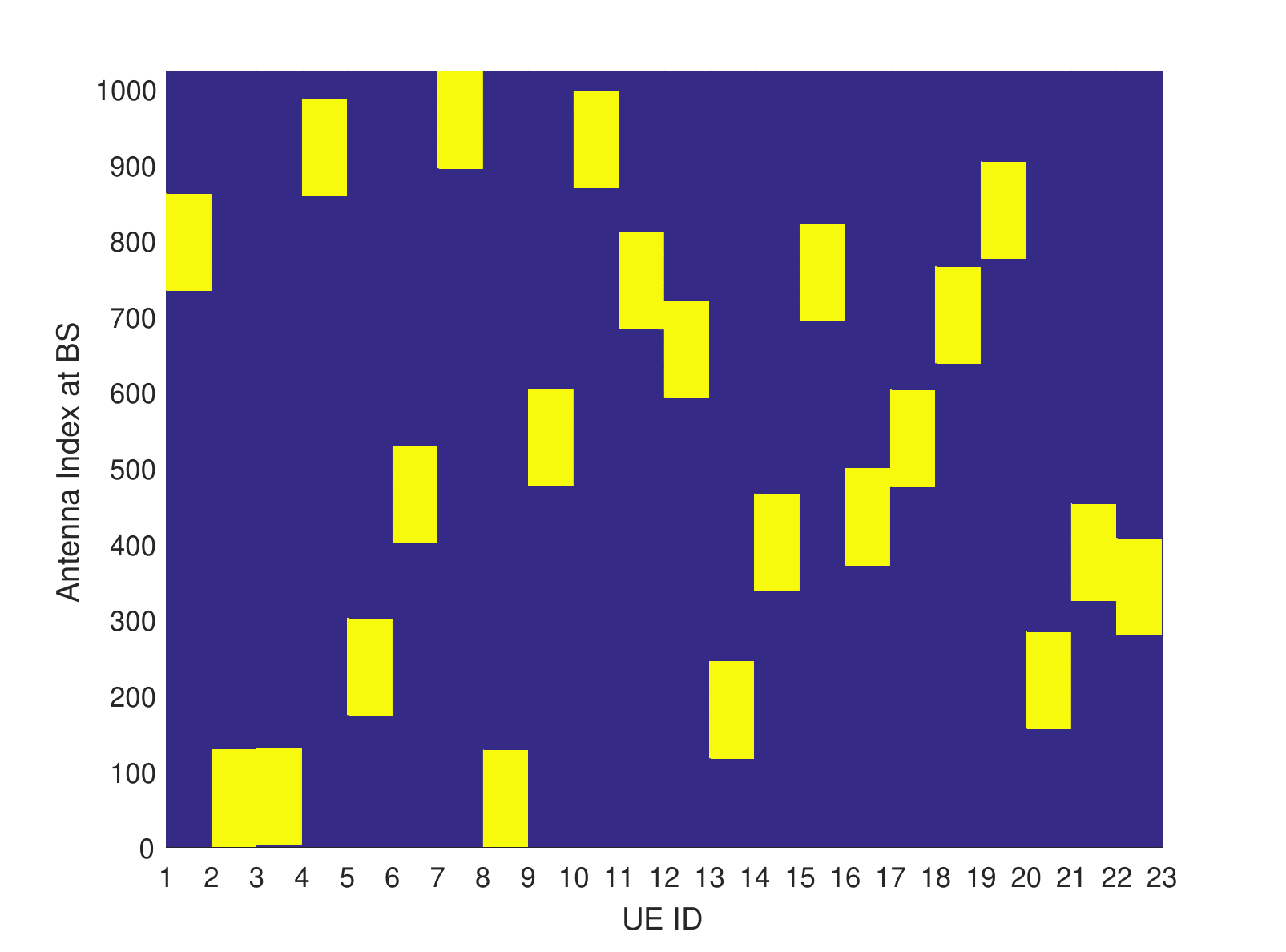}
\caption{ The VR coverage distribution of the total $22$ users in the extra-large scale massive MIMO system.}\label{fig:8}
\end{figure}

Two user scheduling algorithms, i.e., the statistical CSI-based greedy user scheduling algorithm and the statistical CSI-based greedy joint user and subarray algorithm, were proposed in Section V. We firstly investigate the performance of the statistical CSI-based greedy user scheduling algorithm. Fig.\,\ref{fig:6} presents the system sum achievable SEs for the MRC and the LMMSE receivers under the statistical CSI-based greedy user scheduling algorithm.
Users are randomly distributed along the extra-large antenna array as shown in Fig.\,\ref{fig:8} and the number of users to be served is $N_u=12$. The proposed eigenvector-based phase coefficient design is leveraged in the phase shifter-based subarray architecture.

\begin{figure}[h]
\centering
\includegraphics[width=3.6in,height=2.9in]{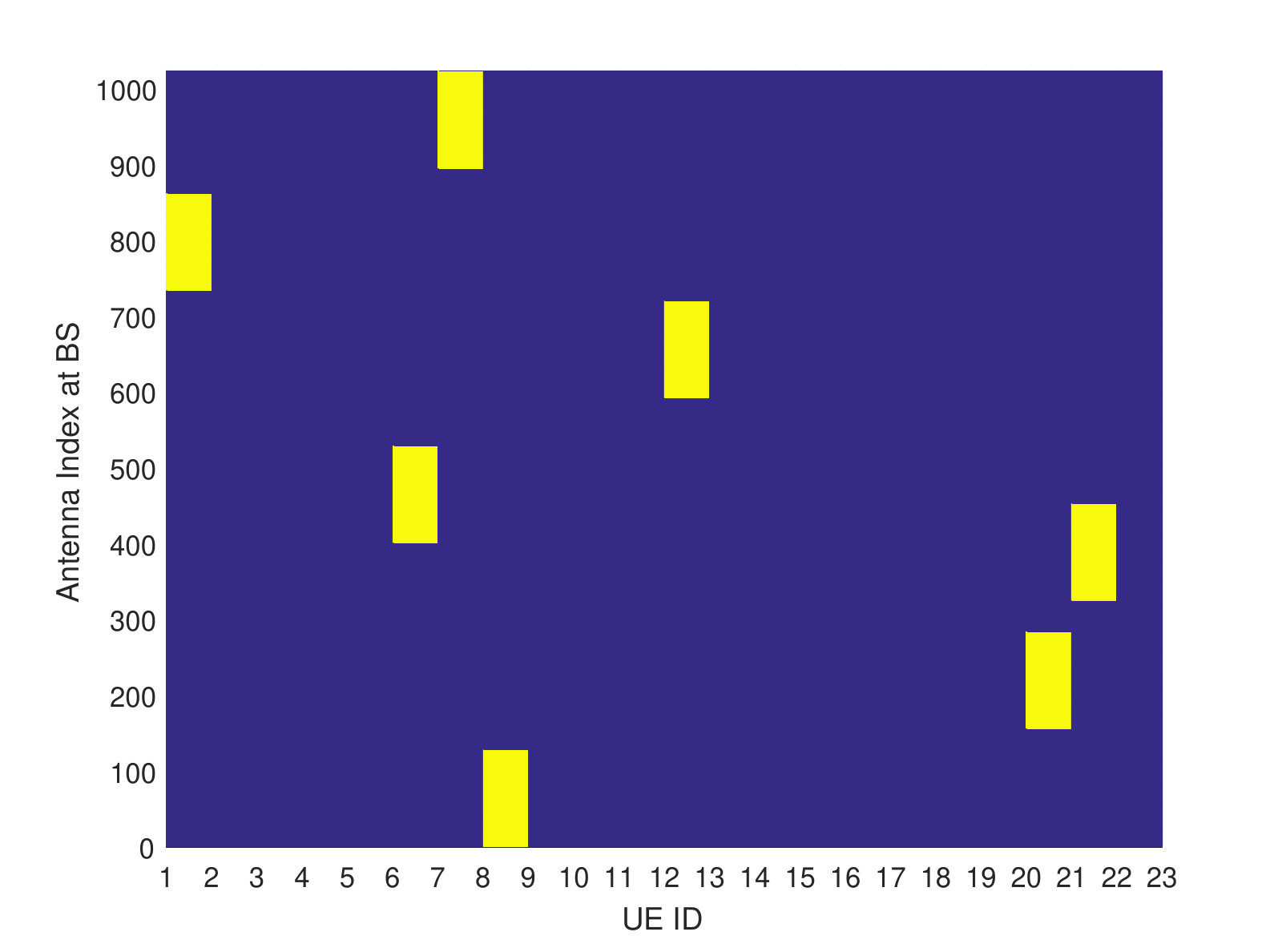}
\caption{ The scheduled $7$ users when using the MRC receiver at the transmit SNR of $36$\,dB under the architecture of the phase shifter-based subarray. The index vector of the scheduled users is $[1,6,7,8,12,20,21]$.}\label{fig:7}
\end{figure}

\begin{figure}[h]
\centering
\includegraphics[width=3.6in,height=2.9in]{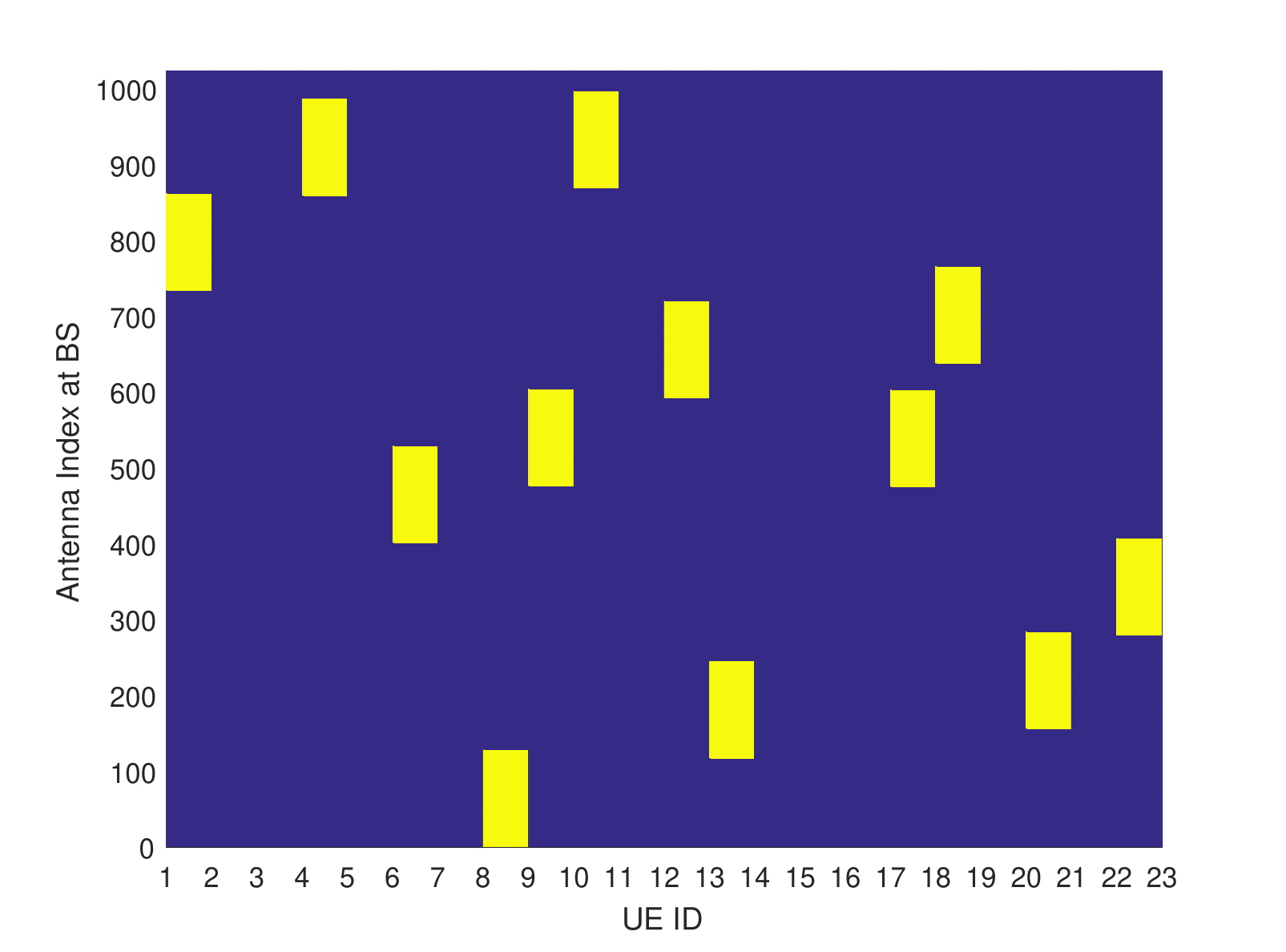}
\caption{ The scheduled $12$ users when using the LMMSE receiver at the transmit SNR of $36$\,dB under the architecture of the phase shifter-based subarray. The index vector of the scheduled users is $[1,4,6,8,9,10,12,13,17,18,20,22]$.}\label{fig:9}
\end{figure}

\begin{figure}[h]
\centering
\includegraphics[width=3.6in,height=2.9in]{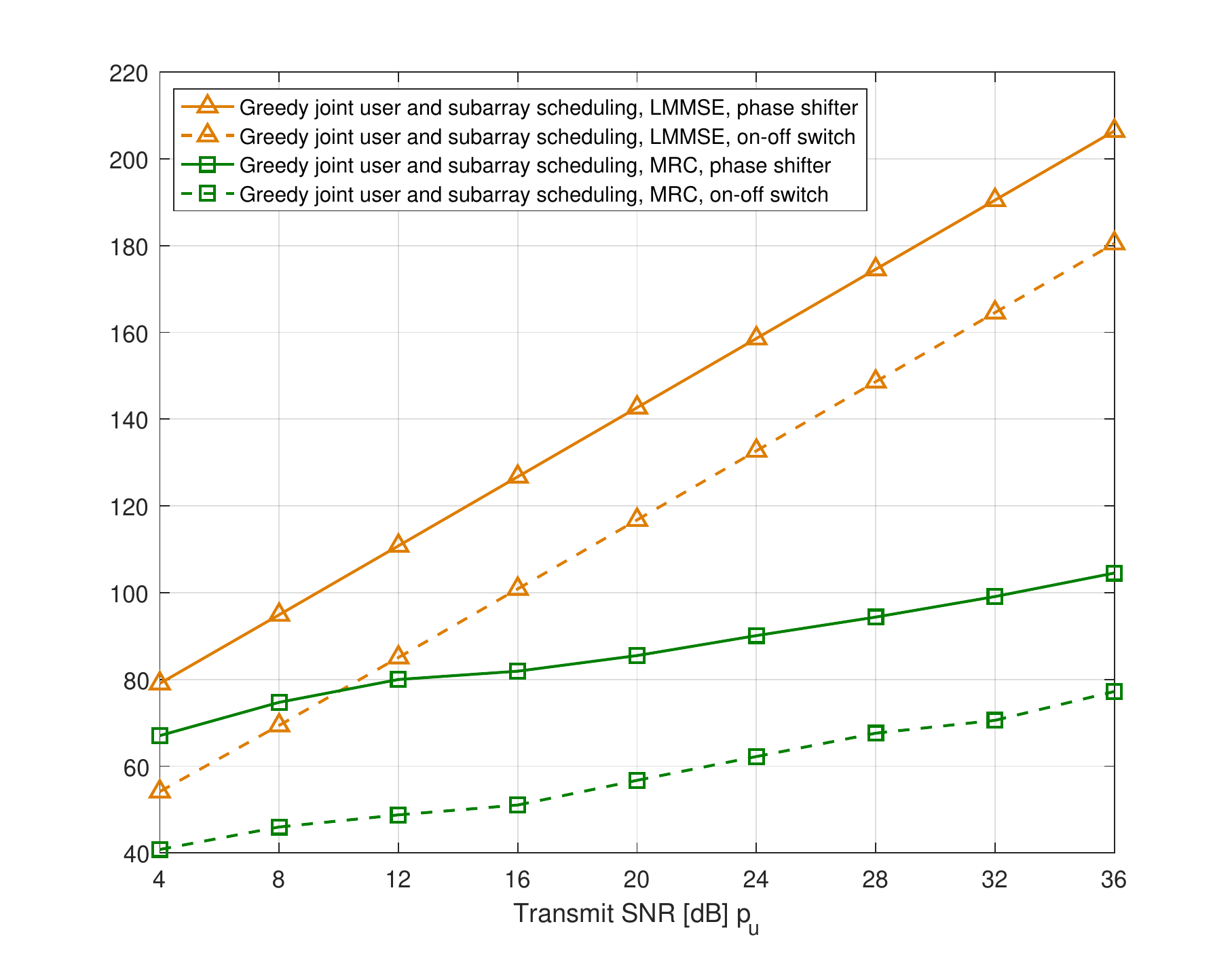}
\caption{ The uplink sum achievable SEs with a phase shifter-based subarray and the on-off switch-based subarray. The parameters $M=1024$, $N=128$, $K=22$, and $E=128$ are selected and users are randomly located along the extra-large antenna array. The statistical CSI-based greedy joint user and subarray scheduling algorithm is utilized. The maximum and minimum number of subarrays that each user can be allocated are $Sub_{max}=8$ and $Sub_{min}=6$ respectively and the number of users to be served is $N_u=12$.}\label{fig:10}
\end{figure}

As can be observed from Fig.\,\ref{fig:6}, regardless of the type of linear receivers (i.e., MRC receiver or LMMSE receiver), the phase shifter-based subarray provides an apparent performance improvement compared to the on-off switch-based subarray; especially with the MRC receiver, nearly $30$\,bits/s/Hz sum achievable SE gains are achieved.
At low SNR, the LMMSE receiver does not apparently outperform the MRC receiver. However, as the transmit SNR increases, inter-user interference becomes stronger owing to the large number of scheduled users and the overlapped VRs. Therefore, the LMMSE receiver begins to exhibit its superiority. Note that nearly $90$\,bits/s/Hz sum achievable SE gains can be acquired by the LMMSE receiver at high SNR.

It is also important to mention that, with an increasing transmit SNR and, thus, stronger inter-user interference, the number of users finally scheduled to be served under the MRC receiver does not always reach the target number of scheduled users, i.e., $N_u$. For example, only $7$ users are scheduled when using the MRC receiver at the transmit SNR of $36$\,dB with a phase shifter-based subarray. The index vector of the finally scheduled users is $[1,6,7,8,12,20,21]$ and Fig.\,\ref{fig:7} plots their positions and corresponding VRs' coverings. Nevertheless, the LMMSE receiver shows its superiority in supporting more users to be served. The scheduled $12$ users at the transmit SNR of $36$\,dB when using the LMMSE receiver are presented in Fig.\,\ref{fig:9}, with their index vector being $[1,4,6,8,9,10,12,13,17,18,20,22]$. Additionally, based on Figs.\,\ref{fig:7} and \ref{fig:9}, it has also been verified that, to maximize the system sum achievable SE, for the MRC receiver, users whose VRs cover different subarrays or those with fewer VR overlaps should be scheduled, while for the LMMSE receiver, users with larger ${\rm{tr}}({\bf{B}}{{\bf{\Theta }}_i})$ should be scheduled as frequently as possible.

Next, we exploit the performance of the statistical CSI-based greedy joint user and subarray scheduling algorithm in Fig.\,\ref{fig:10}.
The maximum and minimum number of subarrays that each user can be allocated to in the joint user and subarray scheduling algorithm are $Sub_{max}=8$ and $Sub_{min}=6$ respectively and we set $N_u=12$. Hence, the number of subarrays (namely RF chains) for each user in the joint user and subarray scheduling algorithm is much less than that in the greedy user scheduling algorithm. Nevertheless, compared with Fig.\,\ref{fig:6}, Fig.\,\ref{fig:10} indicates that the performance of these two linear receivers in the on-off switch-based subarray is only slightly deteriorated, and the performance loss with the phase shifter-based subarray is marginal and even can be neglected. Based on these results, we find that, in extra-large scale massive MIMO systems, it is not necessary to simultaneously turn on all subarrays and RF chains to serve the users. The introduction of dynamic subarray scheduling is beneficial to achieve better system performance with lower system energy consumption.
Besides, the statistical CSI-based greedy joint user and subarray scheduling algorithm collaborating with the on-off switch-based subarray architecture and the LMMSE receiver is a promising practical solution for extra-large scale massive MIMO.

\section{Conclusion}
This paper has investigated the uplink transmission of  extra-large scale massive MIMO systems. In order to perform this task, a subarray-based system architecture was firstly proposed. Then, we derived tight closed-form uplink achievable SE approximations for the extra-large scale massive MIMO system under linear receivers. {Based on these approximations, users with their VRs covering different subarrays or VRs with less overlap should be scheduled simultaneously with a MRC receiver, while users with larger ${\rm{tr}}({\mathbf{B}}{{\mathbf{\Theta }}})$ should be selected with a LMMSE receiver.
In addition, for the subarray with phase shifters, an optimal phase coefficient design was proposed, which relates to the phases of the eigenvectors corresponding to the maximum eigenvalues of the main block matrices of ${{\mathbf{\Theta }}}$.
We also proposed two statistical CSI-based greedy user scheduling algorithms. }
Our numerical results demonstrated that in the extra-large scale massive MIMO system, it is not necessary to simultaneously turn on all subarrays and RF chains to serve the users.
There is a tradeoff between the hardware cost and the system performance. Specifically, the statistical CSI-based greedy joint user and subarray scheduling algorithm collaborating with the on-off switch-based subarray architecture and the LMMSE receiver is a promising practical solution for extra-large scale massive MIMO systems.

\appendices

\section{Proof of Theorem 2}
When the VRs of different user partially overlap, we have
\begin{align}\label{FHF_patially_overlap}
{{\mathbf{F}}^H}{\mathbf{F}} &= {{\mathbf{H}}^H}{\mathbf{BH}} \nonumber \\
           &= \left( {\begin{array}{*{20}{c}}
  {{\mathbf{g}}_1^H{\mathbf{\Theta }}_1^{1/2}{\mathbf{B\Theta }}_1^{1/2}{{\mathbf{g}}_1}}& \ldots &{{\mathbf{g}}_1^H{\mathbf{\Theta }}_1^{1/2}{\mathbf{B\Theta }}_K^{1/2}{{\mathbf{g}}_K}} \\
   \vdots & \ddots & \vdots  \\
  {{\mathbf{g}}_K^H{\mathbf{\Theta }}_K^{1/2}{\mathbf{B\Theta }}_1^{1/2}{{\mathbf{g}}_1}}& \cdots &{{\mathbf{g}}_K^H{\mathbf{\Theta }}_K^{1/2}{\mathbf{B\Theta }}_K^{1/2}{{\mathbf{g}}_K}}
\end{array}} \right).
\end{align}
Since $\mathbb{E}\{ {{\mathbf{g}}_i}{\mathbf{g}}_k^H\}  = {\mathbf{0}},\forall i \ne k$, we obtain $\mathbb{E}\{ {\mathbf{g}}_i^H{\mathbf{\Theta }}_i^{1/2}{\mathbf{B\Theta }}_k^{1/2}{{\mathbf{g}}_k}\}  = {\mathbf{0}},\forall i \ne k$ and
\begin{align}
\mathbb{E}\{ {{\mathbf{I}}_K} + {p_u}{{\mathbf{F}}^H}{\mathbf{F}}\} =&{\rm{diag}}(1 + {p_u}{\mathbf{g}}_1^H{\mathbf{\Theta }}_1^{1/2}{\mathbf{B\Theta }}_1^{1/2}{{\mathbf{g}}_1}, \ldots ,\nonumber\\&\quad\quad\quad 1 + {p_u}{\mathbf{g}}_K^H{\mathbf{\Theta }}_K^{1/2}{\mathbf{B\Theta }}_K^{1/2}{{\mathbf{g}}_K}).
\end{align}
Furthermore, since the VRs of different users only partially overlap, we can safely draw a conclusion that $({{\mathbf{I}}_K} + {p_u}{{\mathbf{F}}^H}{\mathbf{F}})$ is a diagonal-dominant matrix. This diagonal-dominant property has been verified in the numerical results in Section V. Additionally, from (\ref{Rsum_general}) and (\ref{Rk_LMMSE1}), we have
\begin{align}\label{R_patially_overlap}
{R^{\text{LMMSE}}} &= \sum\limits_{i = 1}^K {R_i^{\text{LMMSE}}}  \nonumber \\
           &= \sum\limits_{i = 1}^K {{\mathbb{E}_{\mathbf{h}}}\left\{ {{{\log }_2}\left( {\frac{1}{{{{\left[ {{{\left( {{{\mathbf{I}}_K} + {p_u}{{\mathbf{F}}^H}{\mathbf{F}}} \right)}^{ - 1}}} \right]}_{ii}}}}} \right)} \right\}}  \nonumber \\
          &\mathop  \geqslant \limits^{(a)}  - K{\mathbb{E}_{\mathbf{h}}}\left\{ {{{\log }_2}\left( {\frac{1}{K}\sum\limits_{i = 1}^K {{{\left[ {{{\left( {{{\mathbf{I}}_K} + {p_u}{{\mathbf{F}}^H}{\mathbf{F}}} \right)}^{ - 1}}} \right]}_{ii}}} } \right)} \right\} \nonumber \\
          &\mathop  \geqslant \limits^{(b)}  - K{\log _2}\left( {\frac{1}{K}{\mathbb{E}_{\mathbf{h}}}\left\{ {{\rm{tr}}{{\left( {{{\mathbf{I}}_K} + {p_u}{{\mathbf{F}}^H}{\mathbf{F}}} \right)}^{ - 1}}} \right\}} \right),
\end{align}
where we leverage the inequality of arithmetic and geometric means in (a) and the Jensen's equality in (b). Define ${\mathbf{Z}} \triangleq {{\mathbf{I}}_K} + {p_u}{{\mathbf{F}}^H}{\mathbf{F}}$ and ${\mathbf{\Lambda }} = \rm{diag}(1/{z_{11}},1/{z_{22}}, \ldots ,1/{z_{KK}})$, then ${\mathbf{Z}}$ is a diagonal-dominant matrix. Therefore, according to the Neumann Series \cite{Zhu15On}, for a diagonal-dominant matrix ${\mathbf{Z}}$, its inverse can be expressed as ${{\mathbf{Z}}^{ - 1}} \approx \sum\limits_{n = 0}^L {{{({{\mathbf{I}}_K} - {\mathbf{\Lambda Z}})}^n}{\mathbf{\Lambda }}}$,
where $L$ represents the number of terms used in the Neumann Series. For simplicity, we set $L=1$ and thus
\begin{equation}\label{Z-1}
{{\mathbf{Z}}^{ - 1}} \approx 2{\mathbf{\Lambda }} - {\mathbf{\Lambda Z\Lambda }}.
\end{equation}
Applying (\ref{Z-1}) into (\ref{R_patially_overlap}), we obtain
\begin{align}\label{R_patially_overlap2}
{R^{\text{LMMSE}}} &\geqslant  - K{\log _2}\left( {\frac{1}{K}{\mathbb{E}_{\mathbf{h}}}\left\{ {{\rm{tr}}({{\mathbf{Z}}^{ - 1}})} \right\}} \right) \nonumber \\
            & \approx  - K{\log _2}\left( {\frac{1}{K}{\mathbb{E}_{\mathbf{h}}}\left\{ {{\rm{tr}}\left( {2{\mathbf{\Lambda }} - {\mathbf{\Lambda Z\Lambda }}} \right)} \right\}} \right) \nonumber \\
             &=  - K{\log _2}\left( {\frac{1}{K}{\rm{tr}}\left( {{\mathbb{E}_{\mathbf{h}}}\{ {\mathbf{\Lambda }}\} } \right)} \right).
\end{align}
Moreover, ${z_{ii}} = 1 + {p_u}{\mathbf{g}}_i^H{\mathbf{\Theta }}_i^{1/2}{\mathbf{B\Theta }}_i^{1/2}{{\mathbf{g}}_i}$ and
\begin{align}\label{trGama}
{\rm{tr}}\left( {{\mathbb{E}_{\mathbf{h}}}\{ {\mathbf{\Lambda }}\} } \right) &\geqslant \sum\limits_{i = 1}^K {\frac{1}{{{\mathbb{E}_{\mathbf{g}}}\{ {z_{ii}}\} }}}  \nonumber \\
                    &= \sum\limits_{i = 1}^K {\frac{1}{{1 + {p_u}{\rm{tr}}({\mathbf{B}}{{\mathbf{\Theta }}_i})}}}.
\end{align}
Substituting (\ref{trGama}) into (\ref{R_patially_overlap2}), we have
\begin{align}
{R^{\text{LMMSE}}} &\approx  - K{\log _2}\left( {\frac{1}{K}\sum\limits_{i = 1}^K {\frac{1}{{1 + {p_u}{\rm{tr}}({\mathbf{B}}{{\mathbf{\Theta }}_i})}}} } \right) \nonumber \\
             &\mathop  \leqslant \limits^{(a)} \sum\limits_{i = 1}^K {{{\log }_2}\left[ {1 + {p_u}{\rm{tr}}({\mathbf{B}}{{\mathbf{\Theta }}_i})} \right]},
\end{align}
where (a) utilizes the inequality of arithmetic and geometric means. Hence, the approximated ergodic system sum achievable SE under partially overlapped VR scenario can be expressed as
\begin{equation}\label{R_LMMSE_patially_overlap}
R^{\text{LMMSE,PartialApp}} = \sum\limits_{i = 1}^K{\log _2}\left[ {1 + {p_u}{\rm{tr}}({\mathbf{B}}{{\mathbf{\Theta }}_i})} \right],
\end{equation}
with each user contributing
\begin{equation}\label{Rk_LMMSE_patially_overlap}
R^{\text{LMMSE,PartialApp}}_k = {\log _2}\left[ {1 + {p_u}{\rm{tr}}({\mathbf{B}}{{\mathbf{\Theta }}_k})} \right].
\end{equation}
The proof is concluded.

\emph{Special Scenario:}
When the VRs of different users do not overlap, we have ${{\mathbf{\Theta }}_1} \odot {{\mathbf{\Theta }}_2} \cdots  \odot {{\mathbf{\Theta }}_K} = {\mathbf{0}}$, thus
\begin{align}\label{FHF_donot_overlap}
{{\mathbf{F}}^H}{\mathbf{F}} 
           &= {\rm{diag}}({\mathbf{g}}_1^H{\mathbf{\Theta }}_1^{1/2}{\mathbf{B\Theta }}_1^{1/2}{{\mathbf{g}}_1}, \ldots ,{\mathbf{g}}_K^H{\mathbf{\Theta }}_K^{1/2}{\mathbf{B\Theta }}_K^{1/2}{{\mathbf{g}}_K}),
\end{align}
and
\begin{align}\label{kk_donot_overlap}
\mathbb{E}_{\mathbf{h}}&\left\{ {{{\left[ {{{\left( {{{\mathbf{I}}_K} + {p_u}{{\mathbf{F}}^H}{\mathbf{F}}} \right)}^{ - 1}}} \right]}_{kk}}} \right\} \nonumber \\&= \mathbb{E}_{\mathbf{h}}\{ {(1 + {p_u}{\mathbf{g}}_k^H{\mathbf{\Theta }}_k^{1/2}{\mathbf{B\Theta }}_k^{1/2}{{\mathbf{g}}_k})^{ - 1}}\}  \nonumber \\
                                       &\mathop  \geqslant \limits^{(a)} {(\mathbb{E}_{\mathbf{h}}\{ 1 + {p_u}{\mathbf{g}}_k^H{\mathbf{\Theta }}_k^{1/2}{\mathbf{B\Theta }}_k^{1/2}{{\mathbf{g}}_k}\} )^{ - 1}} \nonumber \\
                                       &\mathop  = \limits^{(b)} {[1 + {p_u}{\rm{tr}}({\mathbf{B}}{{\mathbf{\Theta }}_k})]^{ - 1}},
\end{align}
where (a) applies Jensen's equality $\mathbb{E}\{1/x\}\geqslant 1/\mathbb{E}\{x\}$ for $x>0$ and (b) comes from $\mathbb{E}_{\mathbf{h}}\{ {p_u}{\mathbf{g}}_k^H{\mathbf{\Theta }}_k^{1/2}{\mathbf{B\Theta }}_k^{1/2}{{\mathbf{g}}_k}\}={p_u}{\text{tr}}({\mathbf{B}}{{\mathbf{\Theta }}_k})$. From (\ref{Rk_LMMSE1}), we have
\begin{align}\label{Rk_donot_overlap2}
R_k^{\text{LMMSE}} &= {\mathbb{E}_{\mathbf{h}}}\left\{ {{{\log }_2}\left( {\frac{1}{{{{\left[ {{{\left( {{{\mathbf{I}}_K} + {p_u}{{\mathbf{F}}^H}{\mathbf{F}}} \right)}^{ - 1}}} \right]}_{kk}}}}} \right)} \right\} \nonumber \\
          &\mathop  \geqslant \limits^{(a)} {\log _2}\left( {\frac{1}{{\mathbb{E}_{\mathbf{h}}\left\{ {{{\left[ {{{\left( {{{\mathbf{I}}_K} + {p_u}{{\mathbf{F}}^H}{\mathbf{F}}} \right)}^{ - 1}}} \right]}_{kk}}} \right\}}}} \right),
\end{align}
where Jensen's equality $\mathbb{E}\{\log_2(1/x)\}\geqslant \log_2 (1/\mathbb{E}\{x\})$ for $x>0$ is applied in (a). Combining (\ref{Rk_donot_overlap2}) with (\ref{kk_donot_overlap}), the approximated ergodic achievable SE of the $k$th user with no overlapping VRs can be given by
\begin{equation}\label{Rk_LMMSE_donot_overlap}
R_k^{\text{LMMSE,NoApp}} = {\log _2}\left[ {1 + {p_u}{\rm{tr}}({\mathbf{B}}{{\mathbf{\Theta }}_k})} \right],
\end{equation}
which is consistent with (\ref{Rk_LMMSE_patially_overlap}) as expected.



\ifCLASSOPTIONcaptionsoff
  \newpage
\fi



%

\footnotesize

%








\begin{IEEEbiography}
{\textbf{Xi Yang}} received the B.S.,  M.S. and Ph.D. degrees from Southeast University, Nanjing, China, in 2013, 2016 and 2019, respectively. Her current research interests include wireless communication system prototyping, massive MIMO, and millimeter wave communications.
\end{IEEEbiography}

\begin{IEEEbiography}
{\textbf{Fan Cao}} received the B.S. and M.S. degrees from School of Information Science and Engineering, Southeast University, Nanjing, China in 2017 and 2020, respectively. His main research interests include massive MIMO, metamaterial communication and millimeter wave communications.
\end{IEEEbiography}

\begin{IEEEbiography}
{Michail Matthaiou}(S'05--M'08--SM'13) was born in Thessaloniki, Greece in 1981. He obtained the Diploma degree (5 years) in Electrical and Computer Engineering from the Aristotle University of Thessaloniki, Greece in 2004. He then received the M.Sc. (with distinction) in Communication Systems and Signal Processing from the University of Bristol, U.K. and Ph.D. degrees from the University of Edinburgh, U.K. in 2005 and 2008, respectively. From September 2008 through May 2010, he was with the Institute for Circuit Theory and Signal Processing, Munich University of Technology (TUM), Germany working as a Postdoctoral Research Associate. He is currently a Professor of Communications Engineering and Signal Processing and Deputy Director of the Centre for Wireless Innovation (CWI) at Queen's University Belfast, U.K. after holding an Assistant Professor position at Chalmers University of Technology, Sweden. His research interests span signal processing for wireless communications, massive MIMO systems, hardware-constrained communications, mm-wave systems and deep learning for communications.

Dr. Matthaiou and his coauthors received the IEEE Communications Society (ComSoc) Leonard G. Abraham Prize in 2017. He was awarded the prestigious 2018/2019 Royal Academy of Engineering/The Leverhulme Trust Senior Research Fellowship and also received the 2019 EURASIP Early Career Award. His team was also the Grand Winner of the 2019 Mobile World Congress Challenge. He was the recipient of the 2011 IEEE ComSoc Best Young Researcher Award for the Europe, Middle East and Africa Region and a co-recipient of the 2006 IEEE Communications Chapter Project Prize for the best M.Sc. dissertation in the area of communications. He has co-authored papers that received best paper awards at the 2018 IEEE WCSP and 2014 IEEE ICC and was an Exemplary Reviewer for \textsc{IEEE Communications Letters} for 2010. In 2014, he received the Research Fund for International Young Scientists from the National Natural Science Foundation of China. He is currently the Editor-in-Chief of Elsevier Physical Communication, a Senior Editor for \textsc{IEEE Wireless Communications Letters} and an Associate Editor for the \textsc{IEEE JSAC Series on Machine Learning for Communications and Networks}. In the past, he was an Associate Editor for the \textsc{IEEE Transactions on Communications} and Associate Editor/Senior Editor for \textsc{IEEE Communications Letters}.
\end{IEEEbiography}

\begin{IEEEbiography}
{\textbf{Shi Jin}} (S'06--M'07--SM'17) received the B.S. degree in communications engineering from Guilin University of Electronic Technology, Guilin, China, in 1996, the M.S. degree from Nanjing University of Posts and Telecommunications, Nanjing, China, in 2003, and the Ph.D. degree in information and communications engineering from the Southeast University, Nanjing, in 2007. From June 2007 to October 2009, he was a Research Fellow with the Adastral Park Research Campus, University College London, London, U.K. He is currently with the faculty of the National Mobile Communications Research Laboratory, Southeast University. His research interests include space time wireless communications, random matrix theory, and information theory. He served as an Associate Editor for the IEEE Transactions on Wireless Communications, and IEEE Communications Letters, and IET Communications. Dr. Jin and his co-authors have been awarded the 2011 IEEE Communications Society Stephen O. Rice Prize Paper Award in the field of communication theory and a 2010 Young Author Best Paper Award by the IEEE Signal Processing Society.
\end{IEEEbiography}

\end{document}